\documentclass[twoside,11pt]{article}

\usepackage{graphicx,psfrag,epsf}
\usepackage{enumerate}
\usepackage{natbib}
\usepackage{url} 
\usepackage{bbold}

\usepackage{caption}
\usepackage{subcaption}
\usepackage{amsfonts,amsmath,amsthm,amssymb,color, bbm}
\allowdisplaybreaks[4]

\usepackage{physics}
\usepackage{algorithm}
\usepackage{enumitem}
\usepackage{algorithmic}
\usepackage{caption}
\usepackage{multirow}
\usepackage{enumitem}
\usepackage{booktabs}
\usepackage{mathtools}
\usepackage{setspace}

\usepackage[preprint]{jmlr2e}

\newcommand{\umich}{Department of Statistics\\ University of Michigan\\ Ann Arbor, MI 48109, USA}
\newcommand{\cmu}{Department of Statistics and Data Science\\ Carnegie Mellon University\\ Pittsburgh, PA 15213, USA}

\usepackage{lastpage}
\jmlrheading{}{}{}{}{}{}{Bo Meng, Weijing Tang, Gongjun Xu and Ji Zhu}

\ShortHeadings{Recurrent Event Analysis with ODE}{Meng, Tang, Xu and Zhu}
\firstpageno{1}

\begin{document}

\title{Recurrent Event Analysis with Ordinary Differential Equations}

\author{\name Bo Meng \email bomeng@umich.edu \\
\addr{\umich}
       \AND
       \name Weijing Tang \email weijingt@andrew.cmu.edu\\
       \addr{\cmu}
       \AND
       Gongjun Xu \email gongjun@umich.edu
       \AND Ji Zhu \email jizhu@umich.edu\\
\addr{\umich}
       \AND}

\editor{}

\maketitle

\begin{abstract}
This paper introduces a general framework for analyzing recurrent event data by modeling the conditional mean function of the recurrent event process as the solution to an Ordinary Differential Equation (ODE). This approach not only accommodates a wide range of semi-parametric recurrent event models, including both non-homogeneous Poisson processes (NHPPs) and non-Poisson processes, but also is scalable and easy-to-implement. 
Based on this framework, we propose a Sieve Maximum Pseudo-Likelihood Estimation (SMPLE) method, employing the NHPP as a working model. We establish the consistency and asymptotic normality of the proposed estimator, demonstrating that it achieves semi-parametric efficiency when the NHPP working model is valid. Furthermore, we develop an efficient resampling procedure to estimate the asymptotic covariance matrix. 
To assess the statistical efficiency and computational scalability of the proposed method, we conduct extensive numerical studies, including simulations under various settings and an application to a real-world dataset analyzing risk factors associated with Intensive Care Unit (ICU) readmission frequency.   
\end{abstract}


\section{Introduction}
\label{sec:intro}

Recurrent event data, characterized by repeated occurrences of an event of interest over time, arise in various disciplines, including healthcare, engineering, and other scientific domains.  These events encompass a wide range of phenomena, such as tumor recurrences \citep[][]{krol2016joint}, hospital admissions \citep[][]{thomsen2006methods, vorovich2014biomarker}, infection episodes \citep[][]{grennan2012magnitude}, insurance claims \citep[][]{stahmeyer2019frequency}, and mechanical failures in automotive or aerospace systems \citep[e.g.,][]{cook2007statistical}.  Researchers are often interested in identifying risk factors that influence the frequency of such occurrences.  However, analyzing recurrent event data presents significant challenges, particularly due to censoring mechanisms, such as patient mortality or study termination, which can lead to incomplete observations.

Many statistical models have been developed to analyze censored recurrent event data. One of the most widely used models is the  Cox-type regression model introduced by \cite{andersen1982cox}, which assumes that the underlying recurrent event process is a non-homogeneous Poisson process (NHPP) and that the covariates influence the intensity function multiplicatively. To relax the strict Poisson assumption, researchers such as \cite{pepe1993some}, \cite{lawless1995}, \cite{lin2001transformation}, and \cite{ghosh2002marginal} explored proportional mean and rate models, which assume a multiplicative effect of covariates on the conditional mean function rather than the intensity function.  However, in practice, the assumption of a strict multiplicative effect is often violated, leading to biased results.  Consequently, various semi-parametric models have been proposed as alternatives to the Cox-type model. For instance, \cite{Lin1998AcceleratedFT} and \cite{ghosh2004accelerated} developed the accelerated mean and rate (AM/AR) models, extending the accelerated failure time (AFT) model for survival analysis to recurrent event data. \cite{lin2001semiparametric} introduced a class of semi-parametric transformation models with specified link functions, including Box-Cox and log transformations. \cite{sun2008class} considered a generalized AFT-type model that includes both the AM and proportional mean models as special cases.  Other works have focused on additive models, which allow covariates to have additive effects on an unspecified baseline rate function \citep[][]{lin1995semiparametric, schaubel2006semiparametric, zeng2010semiparametric}. Additionally,  \cite{zeng2006efficient, zeng2007} studied semi-parametric transformation models with random effects, requiring parametric specification of both the transformation function and the distribution of the random effects.

Despite their flexibility, semi-parametric models for recurrent event analysis pose significant computational challenges, particularly in estimating infinite-dimensional functional parameters.
For example, the maximum partial likelihood estimator (MPLE) for the Cox-type model, as described in \cite{andersen1982cox}, is derived by maximizing the partial likelihood function. However, evaluating the partial likelihood for an uncensored subject requires information from subjects who have survived longer, rendering MPLE unsuitable for parallel computing. Moreover, rank-based estimators for AFT-type models, as studied in \cite{Lin1998AcceleratedFT} and \cite{ghosh2004accelerated}, are based on non-smooth objective functions, leading to numerical instability. Nonparametric maximum likelihood estimation (NPMLE) for regression coefficients and cumulative intensity functions in semi-parametric transformation models, introduced by \cite{zeng2006efficient, zeng2007}, is often complicated by the presence of numerous nuisance parameters, particularly in large-scale data scenarios.

Building on the Ordinary Differential Equation (ODE) framework for survival analysis introduced by \cite{tang2023survival}, we propose an ODE-based approach for recurrent event analysis. This framework unifies many existing recurrent event models while providing flexibility to accommodate dependent structures between recurrent events. Specifically, we model the conditional mean function $\mu_x(t)$ of the recurrent event process, which represents the expected number of event occurrences within the interval $[0,t]$ conditional on the covariates $X=x$, as the solution of an ODE. We assume that the derivative of the conditional mean function, $\mu_x'(t)$, is a function of the current time $t$, the covariate $x$, and the current value of the conditional mean function $\mu_x(t)$. This relationship is expressed as
$$\mu_x'(t) = f(t, \mu_x(t), x),$$ where $\mu_x'(t) = d\mu_x(t)/dt$.
By varying the function $f$, our ODE framework can incorporate a wide range of recurrent event models as special cases, including the Cox-type model by \cite{andersen1982cox}, the Accelerated Failure Time (AFT)-type model by \cite{Lin1998AcceleratedFT}, and the semi-parametric transformation model studied by \cite{lin2001transformation} and \cite{zeng2006efficient}.

In contrast to modeling the intensity or hazard function, as in \cite{tang2023survival}, our approach models the conditional mean function, which allows more flexible dependency among recurrent events without imposing assumptions on the underlying counting process. Consequently, our ODE-based framework provides a more general modeling approach that accommodates non-Poisson processes, including non-homogeneous Poisson processes (NHPPs) with random effects, as studied in \cite{nielsen1992counting}, \cite{henderson2000joint}, and \cite{zeng2007}. Since our model does not reply on the Poisson assumption, the maximum likelihood estimation methods in \cite{zeng2007} and \cite{tang2023survival} are no longer applicable.  To address this, we propose a pseudo-likelihood-based estimation method within the ODE framework, which provides new methodological and theoretical contributions beyond the existing literature, as detailed below.

Building on this ODE framework, we propose a gradient-based estimation method that can be applied to a broad class of recurrent event models, offering superior computational efficiency over existing estimation methods. The integration of the ODE framework into recurrent event analysis provides several key advantages. First, the ODE framework enables us to simultaneously evaluate the likelihood functions across different subjects by solving the ODE at different event times. Therefore, the proposed estimation procedure is well-suited for parallel computing, which significantly enhances computational efficiency, particularly when dealing with large-scale datasets.  In contrast, existing approaches such as maximum partial likelihood estimation (MPLE) and rank-based methods cannot access the likelihood functions of all subjects simultaneously, as evaluating the likelihood of an uncensored subject requires information from other subjects. Second, the ODE framework allows us to integrate a spline-based sieve estimator to approximate infinite-dimensional functional parameters. We refer this estimation procedure as the Sieve Maximum Pseudo-Likelihood Estimation (SMPLE) method. Unlike nonparametric maximum likelihood estimation (NPMLE), which requires a large number of nuisance parameters, the SMPLE method can efficiently estimate the functional parameters while substantially reducing computational time in large-scale studies.  In practice, spline-based methods for semi-parametric estimation have been widely studied in survival analysis, recurrent event analysis, and panel count data analysis, as demonstrated by \cite{lu2007estimation, lu2009semiparametric}, \cite{amorim2008regression}, \cite{ding2011sieve}, 
\cite{zeng2007maximum} and \cite{tang2023survival}. Additionally, efficient ODE solvers and polynomial spline estimators have been developed in software such as Python and Matlab. These numerical tools facilitate the evaluation of the objective function and its gradient, enabling efficient maximization of the objective function using existing gradient-based optimization packages. Our method exhibits asymptotically linear time complexity regarding computational efficiency, outperforming many existing estimation approaches under various settings, including rank-based methods in \cite{Lin1998AcceleratedFT} and NPMLEs in \cite{zeng2006efficient, zeng2007}. Notably, while most existing estimation methods are tailored for specific model settings, the proposed SMPLE method offers universal applicability, making it a versatile and computationally efficient alternative.

Moreover, within the ODE framework, we establish the overall convergence rate for the spline-based sieve estimator and the asymptotic normality for the regression coefficients under the general linear transformation setting. While the asymptotic properties for recurrent event analysis and survival analysis have been extensively studied in existing literature, as evident in \cite{andersen1982cox}, \cite{lin2001transformation}, \cite{zeng2007}, \cite{ding2011sieve}, and \cite{tang2023survival}, the introduction of ODEs presents new theoretical challenges for establishing the M-theorem. A key difficulty arises from the bundling of infinite-dimensional nuisance parameters, where an unknown nuisance parameter is expressed as a function of another nuisance parameter and a finite-dimensional Euclidean parameter. This interdependence complicates the maximization of the pseudo-likelihood function.

By contrast, \cite{ding2011sieve} only considers nuisance parameters that are bundled with finite-dimensional Euclidean parameters, making their approach inapplicable to our setting.  Similarly, \cite{tang2023survival} developed an M-theorem for bundle nuisance parameters within an ODE-based framework under the survival analysis setting. However, extending their results to recurrent event models is challenging because the pseudo-likelihood function for recurrent event data can be unbounded, thereby violating a critical assumption in their theorem. To address this issue, we introduce new techniques to control the overall behavior of the pseudo-likelihood function and establish the M-theorem for the proposed spline-based sieve estimator under the recurrent event setting.

Another challenge in developing statistical inference results arises from estimating the covariance matrix of the coefficient estimators, which involves several unknown directional functions. Under general linear transformation models, these functions often lack closed-form solutions, posing computational difficulties. To address this, we derive closed-form solutions for the covariance matrix for many popularly used models, such as Cox-type models, AFT-type models, and linear transformation models with specified transformation functions.  These solutions facilitate the practical estimation of the sieve estimator's covariance matrix and simplify the inference procedure. Additionally, for the general linear transformation model, where closed-form solutions may not be available, we develop an efficient resampling-based method to numerically evaluate the covariance matrix while avoiding computationally intensive density estimation procedures. Our simulation studies validate the theoretical results presented in Section \ref{sec:sec4_thms} and demonstrate the statistical and computational efficiency of our inference method in finite sample settings.

The rest of the paper is organized as follows. In Section~\ref{sec:sec2_setup}, we introduce the ODE framework and describe commonly used recurrent event models covered by our proposed ODE approach. In Section~\ref{sec:sec3_mle}, we present the sieve maximum pseudo-likelihood estimation (SMPLE) method. We establish the asymptotic results in Section~\ref{sec:sec4_thms}. Further, we evaluate the finite-sample performance of our SMPLE method through simulation studies in Section~\ref{sec:5simu}. Finally, we apply our estimation method to a real-world Intensive Care Unit (ICU) dataset in Section~\ref{sec:real data6} and conclude with further discussions in Section 7. All proofs and technical details are provided in the Supplemental Materials.

\section{Model Setup}\label{sec:sec2_setup}
Let $N(t)$ denote the number of recurrent events occurring within the interval $[0, t]$ and let $C$ represent the censoring time, assumed to be independent of event occurrences given the covariates $X\in\mathbf{R}^{d}$. Our goal is to make inferences about the recurrent event process within the given time interval $[0, \tau]$, where the constant $\tau$ denotes the study's duration and is typically determined based on domain knowledge. Note that the recurrent event process $N(\cdot)$ is only observable up to the  time $C$. We define the mean function and rate function of the recurrent event process conditional on covariates $X$ as follows: 
$$\mu_x(t) = E[N(t)|X=x] \mbox{  and  } \mu'_x(t) =E[dN(t)|X=x]/dt.$$ 
The rate function $\mu'_x(t)$ is the dynamic change of the conditional mean function $\mu_x(t)$, quantified by function $f(t,\mu_x(t), x
)$, i.e., $\mu_x(t)$ is the solution to the following ODE:
\begin{equation}
    \left\{ \begin{array}{l}
         \mu'_x(t)=f(t, \mu_x(t), x);  \\
         \mu_x(0)=0. 
    \end{array}\right. \label{ode_intro}
\end{equation}
As discussed in \cite{walter1998first}, under certain smoothness conditions, the ODE (\ref{ode_intro}) has a unique solution. The first line of the ODE specifies how the rate function depends on time $t$, the covariates $x$, and the current value of the conditional mean function $\mu_x(t)$. 
Since $\mu_x(t)$ is an increasing function, $f(\cdot)$ is taken as positive.
Different choices of $f(\cdot)$ lead to different recurrent event models. The second line sets the initial condition of the recurrent event process, and $\mu_x(0) = 0$ indicates that the process begins at time $0$ with no event initially observed. This paper focuses on a general linear transformation model, defined as:
\begin{equation}
f(t, \mu_x(t), x) = \alpha(t)\exp(x^T\beta)q(\mu_x(t)) \label{ode_lt},
\end{equation}
where $\alpha(\cdot)$ and $q(\cdot)$ are positive functions that can be either parametrically specified or left unspecified in a semi-parametric setting, and $\beta = (\beta_1,\ldots, \beta_d)^\top \in\mathbf{R}^{d}$ represents the regression coefficients of covariates $X$. It is important to note that when both $\alpha(\cdot)$ and $q(\cdot)$ are unknown, they may not be identifiable. For ease of reference, we summarize the identifiability conditions under the general linear transformation model in the Supplemental Materials. In practice, setting $\beta_1=1$ and $\alpha\left(t_0\right)=1$ at a fixed time point $t_0\in[0,\tau]$ ensures identifiability of $\alpha(\cdot)$ and $q(\cdot)$. We recommend choosing $t_0$ near the median of all event times to ensure that there are sufficient observations for estimating the unknown functions around $t_0$ and 
within both intervals $[0, t_0]$ and $[t_0, \tau]$.

It is worth noting that the model in (\ref{ode_lt}) is broad enough to include many existing recurrent event models as special cases. For instance, when $q(\cdot) =1$, i.e., 
$$f(t, \mu_x(t), x) = \alpha(t)\exp(x^T\beta),$$
(\ref{ode_lt}) simplifies to the popular proportional rate (Cox-type) model proposed by \cite{andersen1982cox}. Similarly, when $\alpha(\cdot)=1$, i.e., 
$$f(t, \mu_x(t), x) =  q(\mu_x(t))\exp(x^T\beta),$$
it aligns with the accelerated mean (AM/AFT-type) model proposed by \cite{Lin1998AcceleratedFT}. Alternatively, the conditional mean function of the AFT-type model has another format $E[N(t)|X=x] = \mu_0(t\exp(x^T\beta))$ as stated in \cite{Lin1998AcceleratedFT}. In contrast, the ODE framework models the dynamic change of the conditional mean function and provides a more efficient way to estimate the model parameters. Furthermore, when $q(u)$ is a specified positive function, denote $F(t) = \int_0^t 1/q(s)ds$, then we can solve equation (\ref{ode_lt}) and express the conditional mean function as 
$$\mu_x(t)= F^{-1}\left(\int_0^t \alpha(t) \exp\left(x^\top \beta\right)\right).$$ 
This conditional mean function corresponds to the transformation model in \cite{zeng2006efficient} with a specified transformation function. 

Moreover, the proposed ODE framework also accommodates non-Poisson processes. For example, consider a proportional intensity model with random effects and assume the conditional intensity function is $\lambda_x(t|\xi) = \xi\alpha(t)\exp(x^\top \beta)$, where $\xi$ is a random effect variable with expectation equal to $1$. Then the corresponding conditional mean function of this non-Poisson process is 
$$\mu_x(t) = E\Big[\int_0^t \lambda_x(s|\xi)ds \mid X=x\Big] = E[\xi|X=x]e^{x^\top \beta}\int_0^t\alpha(s)ds = e^{x^\top \beta}\int_0^t\alpha(s)ds,$$ 
which satisfies ODE (\ref{ode_lt}) with $q(\cdot)=1$. Such random effect models have been widely explored in the literature \citep[e.g.,][]{nielsen1992counting, oakes1992frailty,liu2004shared, zeng2007, zhao2012analysis}.

\section{Sieve Maximum Pseudo-Likelihood Estimation}\label{sec:sec3_mle}
It is important to note that we only require the conditional mean function $\mu_x(t)$ to satisfy the ODE (\ref{ode_intro}) and make no distributional assumption on the event times. Therefore, without making additional distributional assumptions, the usual maximum likelihood estimation method is not applicable to our framework. For example, to derive the maximum likelihood estimators under the random effect models, it often needs to specify the distribution family of the random effect and assume the conditional recurrent event process to be a Poisson process. However, in our model, the distribution of the underlying random effects can be unspecified, and the recurrent events for each subject can be correlated, which violates the Poisson assumption for the derivation of the exact likelihood function. To address this issue, we employ a pseudo-likelihood estimation approach, which
uses the non-homogeneous Poisson process (NHPP) with the same conditional mean function $\mu_x(t)$ as our working model and obtains the parameter estimators by maximizing the corresponding pseudo-likelihood function. Such maximum pseudo-likelihood estimation approaches have been widely studied under various settings, including survival analysis, recurrent event analysis, and panel count data analysis \citep[e.g.,][]{cox2004note, lu2007estimation,zhao2019nonparametric, sit2021event}. 

We introduce a computationally efficient Sieve Maximum Pseudo-Likelihood Estimation (SMPLE) method and an efficient gradient-based optimization algorithm to estimate the model parameters. Specifically, assume the observed data $\{N_i(t), x_i, c_i, 0\leq t\leq c_i, 1\leq i\leq n\}$ 
are $n$ independently and identically distributed realizations of $\{N(t), X, C, 0\leq t\leq C\}$. Let $\gamma(\cdot ) = \log(\alpha(\cdot))$ and $g(\cdot) = 
\log(q(\cdot))$, where $\alpha(\cdot)$ and 
$q(\cdot)$ are positive functions. 

Next, we calculate the pseudo-likelihood function with the NHPP as our working model. Specifically, in the NHPP setting, the conditional intensity function $\lambda_{x}(t)$ is equal to the conditional rate function $\mu'_{x}(t)$. Therefore,  for individual $i$, the conditional intensity function $\lambda_{x_i}(t)$ is the derivative of the solution of the ODE (\ref{ode_lt}), i.e. $\lambda_{x_i}(t) =\mu'_{x_i}(t) = \exp{x_i^T\beta+\gamma(t)+g(\mu_{x_i}(t))}$, and the loglikelihood function given the observed data under the NHPP setting is
\begin{align}
\ell_n(\beta, \gamma(\cdot), g(\cdot)) =&\frac{1}{n}\sum_{i=1}^n\Bigg\{\int_{0 }^{c_i}\log(\lambda_{x_i}(t))dN_i(t)-\int_0^{c_i}\lambda_{x_i}(t)dt\Bigg\}.\label{log_likelihood}
\end{align}
When the NHPP assumption is violated, equation (\ref{log_likelihood}) no longer represents the true log-likelihood but instead serves as a log pseudo-likelihood function for parameters  $(\beta, \gamma(\cdot), g(\cdot))$. This can be explicitly written as: 
\begin{align}
\begin{aligned}
\ell_{n}(\beta, \gamma(\cdot), g(\cdot))=\frac{1}{n} \sum_{i=1}^{n}\left\{ \int_0^{c_i} \left[ x_i^{T} \beta + \gamma(t)+g({\mu_i(t)})\right]dN_i(t) - \int_0^{c_i}  d\mu_{i}\left(t ; \beta, \gamma, g\right)\right\},
\end{aligned} \label{log-like}
\end{align}
where $\mu_i(t; \beta, \gamma, g)$ is the solution of ODE (\ref{ode_lt}) given covariates $x_i$. Here $\beta$ represents finite-dimensional Euclidean parameters, while $\gamma$ and $g$ are infinite-dimensional functional parameters. 

Directly maximizing (\ref{log-like}) can be computationally challenging due to the infinite-dimensional nuisance parameters $\gamma(\cdot)$ and $g(\cdot)$. To overcome this, we propose to construct a sequence of finite-dimensional sieve spaces as the sample size increases, and these sieve spaces are dense in the original parameter space. Subsequently, we get a sequence of sieve estimators by solving the maximization problem within each sieve space, and the induced finite-dimensional maximization problems are easier to solve than the original problem. The sieve estimation method has been widely explored and shown to be statistically and computationally efficient under various settings \citep[e.g.,][]{shen1994convergence,huang1997sieve, su2012sieve,cao2016sieve, zhao2017sieve,jiang2020reliability,hu2023sieve, shen2023asymptotic}. In practice, we use the polynomial splines to construct the sieve spaces due to their flexibility and computational efficiency in the estimation of complex non-linear functions. 
More implementation details are provided in the following subsections and the numerical comparisons between SMPLE and existing estimation methods are presented in Section \ref{sec:5simu}.

\subsection{Sieve Spaces and Spline-based Semiparametric Estimators}
We start with the construction of spline spaces for the estimation of parameter $\gamma(\cdot)$. Following Definition 4.1 in \cite{schumaker2007spline}, consider a set of distinct and ordered inner knots $\mathcal{T}_{K_n} = \{t_1, \ldots, t_{K_n} \}$ within the interval $[0, \tau]$, dividing it into $K_n+1$ subintervals. For simplicity, let $t_0 = 0$ and $t_{K_n+1} = \tau$. To achieve a desirable convergence rate, we require the number of knots $K_n$ to satisfy $K_n = O(n^{\nu_1})$ and the maximum length of the subintervals to satisfy $\max_{1\leq j \leq K_n+1}|t_j - t_{j-1}| = O(n^{-\nu_1})$ for some $\nu_1 \in (0, 0.5)$. Let $S_n(\mathcal{T}_{K_n}, K_n, p_1)$ denote the space of polynomial splines of order $p_1$ with knots $\mathcal{T}_{K_n}$ as defined in \cite{schumaker2007spline} (see Definition 4.1). According to Corollary 4.10 in \cite{schumaker2007spline}, this class of polynomial splines can be represented by a set of B-spline bases $\{A_j(t), 1\leq j\leq q_n^1\}$, where $q_n^1 = K_n + p_1$. Thus, the spline estimator $\hat{\gamma} \in S_n(\mathcal{T}_{K_n}, K_n, p_1)$ can be expressed as a linear combination of these B-spline base functions, i.e., $\hat{\gamma}(t) = \sum_{j =1}^{q_n^1}a_jA_j(t)$. Consequently, estimating $\gamma$ reduces to estimating the finite-dimensional spline coefficients $\{a_j\}_{j=1}^{q_n^1}$, which is substantially more tractable than directly estimating the infinite-dimensional functional parameter $\gamma$.

Similarly, we construct a spline space for estimating the function $g(\cdot)$. Consider another partition set $\mathcal{T}_{L_n} = \{t_1, \ldots, t_{L_n}\}$ within the interval $[0, \mu]$, where $\mu = \sup_{x\in\mathcal{X}}\mu_0(\tau, x)<\infty$. We assume that $L_n = O(n^{\nu_2})$ and that the maximum subinterval length satisfies $\max_{1\leq j \leq L_n+1}|t_j - t_{j-1}| = O(n^{-\nu_2})$ for some $\nu_2 \in (0, 0.5)$. The polynomial splines of order $p_2$ with knots $\mathcal{T}_{L_n}$, denoted as $S_n(\mathcal{T}_{L_n}, L_n, p_2)$, are similarly represented by B-spline bases $\{B_j(t), 1\leq j\leq q_n^2\}$ with $q^2_n = L_n+p_2$ according to Corollary 4.10 of \cite{schumaker2007spline}, and the spline estimator $\hat{g}\in S_n(\mathcal{T}_{L_n}, L_n, p_2)$ can be expressed as $\hat{g}(t) = \sum_{j = 1}^{q_n^2}b_j B_j(t)$. Substituting $\gamma(\cdot)$ and $g(\cdot)$ in the log pseudo-likelihood function (\ref{log-like}) 
with their spline representations $\hat{\gamma}(\cdot)$ and $\hat{g}(\cdot)$, we obtain the following sieve log pseudo-likelihood function as our objective function:
\begin{equation}
    \ell_{n}(\beta, a, b)=\frac{1}{n} \sum_{i=1}^{n}\left[ \int_0^{c_i} \left\{ x_i^{T} \beta + \sum_{j=1}^{q^1_n}a_j A_j(t)+\sum_{j=1}^{q^2_n}b_j B_j(\mu_i(t;\beta, a, b))\right\}dN_i(t) - \mu_{i}\left(c_i ; \beta, a, b\right)\right],
 \label{sieve-log-like}\end{equation}
where $a = \{a_j\}_{j=1}^{q^1_n}$ and $b = \{b_j\}_{j=1}^{q^2_n}$ are the spline coefficients associated with $\hat{\gamma}(\cdot)$ and $\hat{g}(\cdot)$, respectively. The function $\mu_i(t; \beta, a, b) $ is the solution of the following ODE:
\begin{equation}
  \left\{ \begin{array}{l}
        \mu_{i}^{\prime}(t)=\exp \left(x_i^{T} \beta+\sum_{j=1}^{q^1_n}a_j A_j(t)+\sum_{j=1}^{q^2_n}b_j B_j(\mu_i(t))\right);  \\
        \mu_{i}(0)=0.
   \end{array} \right.
    \label{sieve_ode}
\end{equation}
Maximizing (\ref{sieve-log-like}) yields the parameter estimates $(\hat{\beta}, \hat{a}, \hat{b})$, and the corresponding sieve estimators are given by $\hat{\beta}_n = \hat{\beta}$, $\hat{\gamma}_n(t) = \sum_{j=1}^{q^1_n}\hat{a}_j A_j(t)$, and $\hat{g}_n(t) = \sum_{j=1}^{q^2_n}\hat{b}_j B_j(t)$. It is important to note that maximizing the objective function (\ref{sieve-log-like}) requires evaluation of its gradient. However, the incorporation of the ODE introduces non-trivial challenges for both the evaluation of (\ref{sieve-log-like}) and its gradient for the following reasons. First, the term $\mu_i(c_i;\beta, a, b)$ does not generally have a closed-form expression, as it is defined implicitly through the ODE (\ref{sieve_ode}) given subject-specific covariates $x_i$ and event history. Second, the ODE (\ref{sieve_ode}) differs across individuals due to heterogeneity in covariates $x_i$, and solving these ODEs separately could be computationally intensive. 
To address these challenges, in the next subsection, we introduce a computationally efficient gradient-based optimization algorithm to maximize (\ref{sieve-log-like}).

\subsection{Gradient-based Optimization Algorithm}
To overcome the aforementioned challenges, we adopt local sensitivity analysis from the ODE literature \citep{dickinson1976sensitivity}. Specifically, we construct an auxiliary ODE, whose solution yields the gradient of $\mu_x(t)$ with respect to the model parameters $\theta$, where $\theta = (\beta, a,b)$. Moreover, by applying a specific transformation to the event times for each subject, we are able to solve the given ODE at all event times across different subjects simultaneously, which significantly improves computational efficiency. Specifically, let $t_{ij}$ denote the $j$-th event time for the $i$-th subject, and define the function $$f(t; \mu, \theta) = \exp \left(x^{T} \beta+\sum_{j=1}^{q^1_n}a_j A_j(t)+\sum_{j=1}^{q^2_n}b_j B_j(\mu(t))\right).$$ Given $t_{ij}$, covariates $x_i$, and parameter vector $\theta$, the evaluation of the objective function (\ref{sieve-log-like}) and its gradients involves the following iterative steps:
\begin{enumerate}
\item Solve the ODE (\ref{sieve_ode}) to obtain $\mu_i(t_{ij})$, for $i = 1, \ldots, n$ and $j = 1, \ldots n_i$. Note that we can simultaneously evaluate all the $\mu(t_{ij})$ with only one call of the ODE solver by applying a time transformation on $t_{ij}$. Define $\tilde{t} =e^{x^\top \beta} \int_0^t \exp(\sum_{j=1}^{q^1_n}a_j A_j(s)) ds$, and $\tilde{\mu}(\tilde{t}) = \mu(t)$. Then, the first equation in ODE (\ref{sieve_ode}) can be rewritten into the following formula
\begin{align}
\frac{d\tilde{\mu}(\tilde{t})}{d\tilde{t}} = \exp (\sum_{j=1}^{q^2_n}b_j B_j(\tilde{\mu}(\tilde{t}))),\label{ode_trick}
\end{align} 
which no longer depends on the covariates $x$. Therefore, we can solve the ODE (\ref{ode_trick}) once and evaluate $\mu_i(t_{ij}) = \tilde{\mu}(\tilde{t}_{ij})$, where $\tilde{t}_{ij} = e^{x_i^\top \beta} \int_0^{t_{ij}} \exp(\sum_{j=1}^{q^1_n}a_j A_j(s)) ds$ for $i = 1, \ldots, n$ and $j = 1, \ldots n_i$. This transformation enables a shared computation across all subjects, substantially reducing computational time.

\item 
Evaluate the gradients of the first equation in (\ref{sieve_ode}) with respect to $\theta = (\beta,a,b)$. Let $G(t) = d\mu(t, \theta)/d\theta$ and $\mu'(t) = f(t;\mu, \theta)$. We have the following ODE of $G(t)$:
\begin{align}
\left\{\begin{array}{l}
G^{\prime}(t)=df(t, \mu ; \theta)/d\theta+df(t, \mu ; \theta)/d\mu \cdot G(t);\\
G(0)=0.\label{ode_grad_theta}
\end{array}\right.\notag
\end{align}
By solving this ODE similarly as in part 1, we obtain $G(t_{ij}) = d\mu(t_{ij}, \theta )/d\theta$ for $i = 1, \ldots, n$ and $j = 1, \ldots n_i$.
\end{enumerate}
Therefore, we can numerically evaluate the objective function $\ell_{n}(\theta)$ and its gradient $\ell_{n\theta}'(\theta)$ at each iteration by plugging $\mu(t_{ij})$ 
and $d\mu(t_{ij}, \theta )/d\theta$ into their expressions, respectively. Finally, we can utilize existing gradient-based optimization packages in the software to maximize the objective function (\ref{sieve-log-like}) and obtain the proposed sieve estimator $\hat{\theta}$.

Moreover, we partition the parameter vector $\theta$ into two groups: the regression coefficients $\beta$ and the spline coefficients $a$ and $b$, corresponding to $\hat{\gamma}(t)$ and $\hat{g}(t)$, respectively. When either $\beta$ or $(a, b)$ is held fixed, the objective function is concave with respect to the other. This property motivates the use of a block coordinate ascent algorithm to optimize the objective function (\ref{sieve-log-like}). Specifically, in the $(k+1)$-th iteration, we solve two sub-problems sequentially. First, given the output from the $k$-th iteration, we update $(a^{(k+1)}, b^{(k+1)})$ by maximizing $\ell_n(\beta^{(k)}, a, b)$ subject to the identifiability constraint $\hat{\gamma}_n(t_0) = \sum_{j=1}^{q_1}a_j A_j(t_0) = 0$, where $t_0$ is a pre-specified time point. Second, we update $\beta^{(k+1)}$ by maximizing $\ell_n(\beta, a^{(k+1)}, b^{(k+1)})$ subject to the constraint $\beta_1^{(k+1)}=1$ for identifiability. These sub-problems can be efficiently solved using standard computational packages. In our implementation, we leverage Matlab's ``Optimization Toolbox'', utilizing ``fminunc'' for unconstrained optimization and ``fmincon'' for constrained problems. For parameter initialization, we adopt random initialization for the Cox-type model due to the global concavity of the objective function. For non-concave models, such as the general linear transformation model, we initialize $\beta$ using the estimated $\hat{\beta}$ obtained from the Cox-type model and set the initial values of the spline coefficients $a$ and $b$ to zero. Our simulation studies demonstrate the robustness and efficiency of the proposed gradient-based optimization methods across a variety of model settings.

\section{Asymptotic Results}\label{sec:sec4_thms}
In this section, we establish the theoretical properties of the proposed Sieve Maximum Pseudo-Likelihood Estimator (SMPLE). As noted in Section \ref{sec:intro}, due to the unbounded nature of the pseudo-likelihood function and the bundled infinite-dimensional nuisance parameters within our framework, the M-estimation theorems in \cite{ding2011sieve} and \cite{tang2023survival} are not applicable to our setting. To address these challenges, we develop new techniques to control the overall behavior of the log pseudo-likelihood function and establish the consistency, convergence rate, and asymptotic normality of the proposed estimator. To ensure identifiability when both $\gamma(\cdot)$ and $g(\cdot)$ are unspecified, we assume $\beta\in \mathbf{R}^{d+1}$ with $\beta_1 = 1$, and $\gamma(t_0) = 0$, where $t_0$ is a pre-specified time point.  Moreover, define $Pf = \int f(x) Pr(dx)$ where $Pr$ is a probability measure $Pr$. In what follows, we will use $\beta \in \mathbf{R}^d$ to denote the unknown part of the regression parameter, excluding the fixed first component.

We start with introducing the following regularity conditions.

\begin{itemize}
\item[(C1)] The true parameter $\beta_{0}$ is an interior point of a compact set $\mathcal{B} \subset \mathbf{R}^{d}$.
    
\item[(C2)] The domain $\mathcal{X}$ of the covariates $X$ is a compact subset of $\mathbf{R}^{d+1}$, the density of $X$ is  bounded below by a positive constant $c>0$, and the matrix $P(XX^\top)$ is non-singular. 
    
\item[(C3)] There exists a finite truncation time $\tau_0>0$ and a constant $\delta_{0}>0$ such that, $Pr(C>\tau_0 \mid X) \geq \delta_{0}>0$ almost surely with respect to the distribution of $X$, where $C$ denotes the censoring time. 
    
\item[(C4)]  Consider a class of functions $f$ that are bounded and have bounded derivatives up to order $k$. Additionally, the $k$-th derivative $f^{(k)}$ satisfies the $m$-Hölder continuity condition, i.e., for any $x$, $y\in [a, b]$,
$$
|f^{(k)}(x)-f^{(k)}(y)| \leq L|x-y|^{m},
$$
where $m \in(0,1]$, and $L<\infty$ is a constant. Define $p=m+k$ and denote the set of these functions as $S^p([a, b])$. 
We assume the true parameter functions satisfy $\gamma_{0}\in \Gamma^{p_{1}} := \left\{ \gamma \in S^{p_{1}}([0, \tau]): \gamma(t_0)=0 \right\}$ with $p_{1} \geq 2$, and $g_{0}\in \mathcal{G}^{p_{2}} := S^{p_{2}}\left(\left[0, \mu+\delta_{1}\right]\right)$ with $p_{2} \geq 3$ and $\delta_{1}>0$. The time point $t_0$ is pre-specified in the identifiability constraint imposed earlier.

\item[(C5)] Define $R(t)=\int_{0}^{t} \exp \left(\gamma_{0}(s)\right) d s, V= X^{T} \beta_{0}$, and $U(t)=e^{V} R(t)$. There exists a constant $\eta_{1} \in(0,1)$ such that for all $u \in \mathbf{R}^{d}$ with $\|u\|=1$, the following inequality:
$$
u^{T} \operatorname{Var}\left(X \mid U(t), V\right) u \geq \eta_{1} u^{T} P\left(X X^{T} \mid U(t), V\right) u ,
$$ 
holds for any $t \in [0, \tau]$, almost surely.
\end{itemize}
Conditions (C1)-(C3) are standard assumptions in the analysis of recurrent event data. Condition (C4) imposes smoothness requirements on the functional parameters $\gamma_0$ and $g_0$, which are common in spline-based semi-parametric estimation (see, e.g., \cite{ding2011sieve}). Condition (C5) 
was introduced by \cite{wellner2007two} in the context of panel count data and subsequently employed in \cite{ding2011sieve} for the transformation model with a known transformation function. 

We now introduce additional notation to facilitate the investigation of the theoretical properties of the proposed sieve estimator. Let $\mu(t, x, \beta, \gamma, g)$ denote the solution to the following ODE:
\begin{equation}
\begin{aligned}
\left\{\begin{array}{l}
\mu^{\prime}(t)=\exp \left(x^{T} \beta+\gamma(t)+g(\mu(t))\right); \\
\mu(0)=0,
\end{array}\right.
\end{aligned} \label{ode_lt_thm}
\end{equation}
and define the composite function $g(\mu(t, x, \beta, \gamma, g))$, mapping from $\mathcal{T}\times\mathcal{X}\times\mathcal{B}\times \Gamma^{p_1}$ into $\mathbf{R}$. We define the class of functions
\begin{align*}
\mathcal{H}^{p_{2}}=\Big\{ \zeta(\cdot, \beta, \gamma): \zeta(t, x, \beta, \gamma)=& g(\mu(t, x, \beta, \gamma, g)), t \in[0, \tau], x \in \mathcal{X}, \beta \in \mathcal{B}, \gamma \in \Gamma^{p_{1}}, \\
&g \in \mathcal{G}^{p_{2}} \text { with } \sup _{t \in[0, \tau], x \in \mathcal{X}}|\mu(t, x, \beta, \gamma, g)| \leq \mu \Big\}.
\end{align*} 
The $L_2$ norm of a function $\zeta(\cdot, \beta, \gamma)\in \mathcal{H}^{p_2}$ is defined as  
\[\|\zeta(\cdot, \beta, \gamma)\|_{2}=\left[\int_{\mathcal{X}} \int_{0}^{\tau}\zeta(t, x, \beta, \gamma)^{2} d \mu_{0}(t, x) d F_{X}(x)\right]^{1 / 2},\]
where $F_X(x)$ is the cumulative distribution function of covariates $X$. Let $\theta = (\beta, \gamma(\cdot), \zeta(\cdot, \beta, \gamma))$ denote the collection of parameters, with the corresponding parameter space  $\Theta = \mathcal{B}\times\Gamma^{p_1}\times\mathcal{H}^{p_2}$. The true parameter is denoted by $\theta_0 = (\beta_0, \gamma_0(\cdot), \zeta_0(\cdot, \beta_0, \gamma_0))$, where $\zeta_0(t) = g_0(\mu(t, x, \beta_0, \gamma_0, g_0))$. The distance between two parameter values $\theta_1$, $\theta_2 \in \Theta$ is defined as
\[
d\left(\theta_{1}, \theta_{2}\right)=\left(\left\|\beta_{1}-\beta_{2}\right\|^{2}+\left\|\gamma_{1}-\gamma_{2}\right\|_{2}^{2}+\left\|\zeta_{1}\left(\cdot, \beta_{1}, \gamma_{1}\right)-\zeta_{2}\left(\cdot, \beta_{2}, \gamma_{2}\right)\right\|_{2}^{2}\right)^{1 / 2},
\]
where $\|\cdot\|$ denotes the Euclidean norm and $\|\cdot\|_2$ the $L_2$ norm. We further define the following collections of functions \[\Gamma_n^{p_1} = \left\{\gamma \in S_{n}\left(T_{K_{n}^{1}}, K_{n}^{1}, p_{1}\right): \gamma(t_0)=0\right\},\  \mathcal{G}_{n}^{p_{2}}=S_{n}\left(T_{K_{n}^{2}}, K_{n}^{2}, p_{2}\right),\] 
\[
\mathcal{H}_{n}^{p_{2}}=\left\{\zeta(\cdot, \beta, \gamma): \zeta(t, x, \beta, \gamma)=g(\mu(t, x, \beta, \gamma, g)), g \in \mathcal{G}_{n}^{p_{2}}, t \in[0, \tau], x \in \mathcal{X}, \beta \in \mathcal{B}, \gamma \in \Gamma_{n}^{p_{1}}\right\},
\] 
and define the sieve space as $\Theta_n = \mathcal{B}_n\times \Gamma_n^{p_1}\times \mathcal{H}_n^{p_2}$. By construction, we have $\Theta_n\subset\Theta_{n+1}\subset\cdots\subset\Theta$ for $n\in \mathbf{N}$. 
With the above notation, we now introduce the additional regularity conditions:
\begin{itemize}
\item[(C6)] Let $\psi(t, x, \beta, \gamma, g)=x^{T} \beta+\gamma(t)+g(\mu(t, x, \beta, \gamma, g))$, and denote its functional derivatives with respect to $\gamma(\cdot)$ and $g(\cdot)$, along directions $v(\cdot)$ and $w(\cdot)$ at $\theta_0$, by $\psi_{0 \gamma}^{\prime}(t, x)[v]$ and $\psi_{0 g}^{\prime}(t, x)[w]$, respectively. (Formal  definitions are provided in the proof of Lemma 2 in the appendix.) Then for any $v \in \Gamma^{p_{1}}$ and $w \in \mathcal{G}^{p_{2}}$, there exists a constant $0<\eta_{2}<1$ such that
\begin{align*}
&\left(P\left\{\int_0^C \psi'_{0\gamma}(t, x)[v] \psi'_{0g}(t, x)[w]d\mu_0(t, x)\right\}\right)^2\\
&\quad \quad \leq\eta_2 P\left\{\int_0^C \psi'_{0\gamma}(t, x)[v]^2d\mu_0(t, x)\right\} P\left\{\int_0^C\psi'_{0g}(t, x)[w]^2d\mu_0(t, x)\right\},
\end{align*}
almost surely.
\item[(C7)] There exists a positive finite constant $\omega_0$ such that for any $x\in \mathcal{X}$, the conditional expectation $P\{e^{\omega_0N(\tau)}N(t)|X=x\}$ is a bounded and Lipschitz continuous function of $t \in [0, \tau]$, with Lipschitz constant $L<\infty$.
\end{itemize}
Condition (C6) ensures no strong collinearity between the functional derivatives with respect to $\gamma$ and $g$, which is essential for establishing consistency and calculating asymptotic variance of the estimator.
Condition (C7) is a technical requirement that controls the growth of the recurrent event process, and it holds, for example, when the true underlying process follows a non-homogeneous Poisson process (NHPP) or an NHPP with light-tailed random effects. Similar assumptions have been used in related work such as  \cite{lu2007estimation, lu2009semiparametric}. 

Let $\hat{\theta}_n = (\hat{\beta}_n, \hat{\gamma}_n(\cdot), \hat{\zeta}_n(\cdot, \hat{\beta}_n, \hat{\gamma}_n) )$ denote the maximizer of the objective function (\ref{sieve-log-like}) within each sieve space, where $\hat{\zeta}_n(\cdot, \hat{\beta}_n, \hat{\gamma}_n) = \hat{g}(\mu(t, x, \hat{\beta}_n, \hat{\gamma}_n,\hat{g}_n)$. We first establish the consistency of $\hat{\theta}_n$ in the following theorem.
\begin{theorem}[Consistency]\label{thm1}
Suppose conditions (C1)-(C7) hold with $\omega_0>\max\{(d+1)M_\beta M_X+M_\gamma+M_\zeta, 1\}$. Then, we have 
\[
d(\hat{\theta}_n, \theta_0)\rightarrow 0,
\] 
in probability as $n\rightarrow \infty$.
\end{theorem} 
The quantities $M_\beta$, $M_X$, $M_\gamma$, and $M_\zeta$ are rigorously defined in the Supplemental Materials. Theorem \ref{thm1} establishes the consistency of the proposed sieve estimator $\hat{\theta}_n$, which provides the first step for analyzing its convergence rate. Notably, our proof technique differs from related studies such as \cite{ding2011sieve} and \cite{tang2023survival}. 
Specifically, we leverage the general M-estimation framework developed in \cite{chen2007large}, rather than relying on the convergence theorem of \cite{shen1994convergence}, which was used in those earlier works for survival models.
A key technical challenge arises from the unbounded nature of the pseudo-likelihood function in the recurrent event setting, which makes it difficult to uniformly control the maximum norm of the objective function---an essential step in the proof strategy of \cite{ding2011sieve} and \cite{tang2023survival}. To address this issue, we employ the Bernstein norm in place of the maximum norm, allowing us to control the expected behavior of the pseudo-likelihood function. We then derive a Bernstein-type upper bound analogous to those obtained in previous studies.
This adaptation enables us to apply Lemma 3.4.3 of \cite{vaart1997weak} and Theorem 3.1 of \cite{chen2007large} to rigorously establish Theorem~\ref{thm1}.

Next, we establish the convergence rate of the proposed sieve estimator in the following theorem.
\begin{theorem}[Convergence rate of $\hat{\theta}_n$]\label{thm2}
Suppose $\nu_1$ and $\nu_2$ satisfy $\max \left\{\frac{1}{2\left(2+p_{1}\right)}, \frac{1}{2 p_{1}}-\frac{\nu_{2}}{p_{1}}\right\}<\nu_{1}<\frac{1}{2 p_{1}}$,  $\max \left\{\frac{1}{2\left(1+p_{2}\right)}, \frac{1}{2\left(p_{2}-1\right)}-\frac{2 \nu_{1}}{p_{2}-1}\right\}$ $<\nu_{2}<\frac{1}{2 p_{2}}$, and  $2 \min \left\{2 \nu_{1}, \nu_{2}\right\}>\max \left\{\nu_{1}, \nu_{2}\right\}$, and the conditions in Theorem \ref{thm1} hold. Then, we have 
\[
d(\hat{\theta}_n, \theta_0) = O_p(n^{-\min\{ p_1\nu_1, p_2\nu_2, \frac{1-\max\{\nu_1, \nu_2\}}{2}\}}).
\]
\end{theorem}

Note that $p_1$ and $\nu_1$ in Theorem \ref{thm2} correspond to the spline space for $\hat{\gamma}_n$, and $p_2$ and $\nu_2$ correspond to the spline space for $\hat{g}_n$. The regularity conditions on $\nu_1$ and $\nu_2$ are satisfied when $p_1$ and $p_2$ are not far from each other. Moreover, when $p_1 = p_2 = p$ and $\nu_1 = \nu_2 = \nu = \frac{1}{1+2p}$, the proposed estimator achieves the optimal convergence rate $d(\hat{\theta}_n, \theta_0) = O_p(n^{-\frac{p}{1+2p}})$ under the nonparametric regression setting. Although the overall convergence rate of the sieve estimator $\hat{\theta}_n$ is slower than $n^{-\frac{1}{2}}$, it can be shown that the estimator $\hat{\beta}_n$ for the regression coefficients is still asymptotically normal and achieves the semi-parametric efficiency bound when the underlying true recurrent event process follows a non-homogeneous Poisson process (NHPP). 

To establish the asymptotic normality of $\hat{\beta}$, we introduce two additional regularity conditions.
\begin{itemize}
\item[(C8)] There exist $\mathbf{v}^{*}=\left(v_{1}^{*}, \cdots, v_{d}^{*}\right)^{T}$ and $\mathbf{w}^{*}=\left(w_{1}^{*}, \cdots, w_{d}^{*}\right)^{T}$, with $v_{j}^{*} \in \Gamma^{2}$ and $w_{j}^{*} \in \mathcal{G}^{2}$ for $j=1, \ldots, d$, such that for any $v\in \Gamma^{p_1}$ and $w\in \mathcal{G}^{p_2}$, the following orthogonality conditions hold:
\begin{align}
P\left\{ \int_0^C\mathbf{A}^{*}(U(t), X) \psi_{0 \gamma}^{\prime}(t, X)[v]d\mu_0(t)\right\}=0,\label{condition_8_1}
\end{align} and 
\begin{align}
P\left\{\int_0^C \mathbf{A}^{*}(U(t), X) \psi_{0 g}^{\prime}(t, X)[w]d\mu_0(t)\right\}=0,\label{condition_8_2}
\end{align} 
where $U(t)$ is defined in condition (C5), $\psi_{0 \gamma}^{\prime}(t, X)[v]$ and $\psi_{0 g}^{\prime}(t, X)[w]$ are the functional derivatives defined in Condition (C6). In equations (\ref{condition_8_1}) and (\ref{condition_8_2}), $\mathbf{A}^*(U(t), X)$ represents the function $\mathbf{A}^*(t, X)$ evaluated with $U(t)$ substituted for the time parameter and $\mathbf{A}^*(t, X)$ is explicitly defined as:
$$
\begin{aligned}
\mathbf{A}^*(t, X)= & \left(g_0^{\prime}\left(\tilde{\mu}_0(t)\right) \exp \left(g_0\left(\tilde{\mu}_0(t)\right)\right) t+1\right) X \\
& -g_0^{\prime}\left(\tilde{\mu}_0(t)\right) \exp \left(g_0\left(\tilde{\mu}_0(t)\right)\right) \int_0^t \mathbf{v}^*\left(R^{-1}\left(s e^{-V}\right)\right) d s+\mathbf{v}^*\left(R^{-1}\left(t e^{-V}\right)\right) \\
& -g_0^{\prime}\left(\tilde{\mu}_0(t)\right) \exp \left(g_0\left(\tilde{\mu}_0(t)\right)\right) \int_0^{\tilde{\mu}_0(t)} \exp \left(-g_0(s)\right) \mathbf{w}^*(s) d s+\mathbf{w}^*\left(\tilde{\mu}_0(t)\right),
\end{aligned}
$$
where $\tilde{\mu}_0(t)$ satisfies the differential equation $\tilde{\mu}_0^{\prime}(t)=\exp \left(g_0\left(\tilde{\mu}_0\right)\right)$ with initial condition $\tilde{\mu}_0(0)=0$.
\item[(C9)] The matrix
\[ 
A = P\left\{\mathbf{A}^{*}(U(t), x)^{\otimes 2}d\mu_0(t, x)\right\},
\] 
is non-singular, where $a^{\otimes 2}=a a^{T}$ for any vector $a$.
\end{itemize}
Condition (C8) requires the existence of  least favorable directions $\mathbf{v}^{}$ and $\mathbf{w}^{}$, which are used to construct the asymptotic covariance matrix of $\hat{\beta}$ in Theorem \ref{thm3}.  Closed-form solutions for $\mathbf{v}^{}$ and $\mathbf{w}^{}$ under the Cox-type model and the linear transformation model with a known transformation function $\gamma_0$ are provided in the Supplemental Materials. Note that, when the underlying true recurrent event process follows a non-homogeneous Poisson process, $M(t)$ represents the martingale of this recurrent event process, and $A$ becomes the Fisher information matrix $I(\beta_0)$ for the regression coefficient $\beta_0$. Thus, Condition (C9) is a natural extension of the standard non-singularity condition on the Fisher information matrix, which is commonly assumed in recurrent event analysis.

With the addition of the regularity conditions (C8) and (C9), we now establish the asymptotic normality and semi-parametric efficiency of the estimator $\hat{\beta}_n$.
\begin{theorem}[Asymptotic normality of $\hat{\beta}_n$]\label{thm3}
Suppose conditions in Theorem 2 and (C8)--(C9) hold, and $\omega_0>\max\{M_v+M_{\zeta_{\gamma}}, M_h, M_X + M_{\zeta_{\beta}}\}$. Then, we have 
\[
\sqrt{n}\left(\hat{\beta}_{n}-\beta_{0}\right)=\sqrt{n} A^{-1} P_{n} \Xi^{*}\left(\beta_{0}, \gamma_{0}, \zeta_{0} ; W\right)+o_{p}(1) \rightarrow_{d} N\left(0, A^{-1} B A^{-T}\right)
\]
where the matrix $A$ is defined in Condition (C9), and 
$$
\begin{aligned}
\Xi^{*}\left(\beta_{0}, \gamma_{0}(\cdot), \zeta_{0}\left(\cdot, \beta_{0}, \gamma_{0}\right) ; W\right)=& \ell_{\beta}^{\prime}\left(\beta_{0}, \gamma_{0}(\cdot), \zeta_{0}\left(\cdot, \beta_{0}, \gamma_{0}\right) ; W\right)\\
&-\ell_{\gamma}^{\prime}\left(\beta_{0}, \gamma_{0}(\cdot), \zeta_{0}\left(\cdot, \beta_{0}, \gamma_{0}\right) ; W\right)\left[\boldsymbol{v}^{*}\right] \\
&-\ell_{\zeta}^{\prime}\left(\beta_{0}, \gamma_{0}(\cdot), \zeta_{0}\left(\cdot, \beta_{0}, \gamma_{0}\right) ; W\right)\left[\boldsymbol{h}^{*}\left(\cdot, \beta_{0}, \gamma_{0}\right)\right],
\end{aligned}$$
and
$$
\begin{aligned}
B=& P\left\{\Xi^{*}\left(\beta_{0}, \gamma_{0}(\cdot), \zeta_{0}\left(\cdot, \beta_{0}, \gamma_{0}\right) ; W\right) \Xi^{*}\left(\beta_{0}, \gamma_{0}(\cdot), \zeta_{0}\left(\cdot, \beta_{0}, \gamma_{0}\right) ; W\right)^{T}\right\},
\end{aligned}
$$where $\mathbf{v}^*$ and $\mathbf{w}^*$ are functional vectors defined in condition (C8).
\end{theorem}
Rigorous definitions of $\ell_{\beta}'(\theta_0; W)$, $\ell_{\gamma}^{\prime}(\theta_0; W)\left[\boldsymbol{v}^{*}\right]$, $\ell_{\zeta}^{\prime}(\theta_0; W)\left[\boldsymbol{h}^{*}\left(\cdot, \beta_{0}, \gamma_{0}\right)\right]$, $M_v$, $M_{\zeta_{\gamma}}$, $M_h$, $M_X$ and $M_{\zeta_{\beta}}$ are given in the Supplemental Materials.
Note that, if the true recurrent event process follows a non-homogeneous Poisson process (NHPP), we have $A = B = I(\beta_0)$, where $I(\beta_0)$ is the Fisher information matrix for $\beta_0$. Consequently, the asymptotic covariance matrix simplifies to $\Sigma = A^{-1}BA^{-T}
= I^{-1}(\beta_0)$, indicating that $\hat{\beta}_n$ achieves the semiparametric efficiency bound under the NHPP setting. In practice, matrix $B$ can be estimated using the empirical information matrix. However, obtaining a closed-form expression for matrix $A$ is challenging, as the least favorable directions $\mathbf{v}^{*}$ and $\mathbf{w}^{*}$ are implicitly defined through the solution of two additional equations, details of which are discussed in the Supplemental Materials. 
 
To address this issue, we derive closed-form solutions for $\mathbf{v}^{*}$ and $\mathbf{w}^{*}$ under commonly used models, such as the Cox-type model and the linear transformation model with a specified transformation function. All technical details are provided in the Supplemental Materials. For the general transformation model with unspecified $\gamma(\cdot)$ and $g(\cdot)$, we adopt the resampling method presented in \cite{resampling} to estimate the matrix $A$, leveraging the sandwich formula $A^{-1}BA^{-T}$
for the asymptotic covariance matrix, which aligns with the formulation in \cite{resampling}. Specifically, let $U_n(\theta)= {\partial l_n(\theta)}/{\partial \theta}$, and let $Z$ be a mean-zero random vector independent of the observed data. Recall that the estimator $\hat{\theta}_n$ satisfies the estimating equation $U_n(\theta) = 0$ and converges to $\theta_0$ in probability (as shown in Theorem \ref{thm1}). Therefore, for sufficiently large $n$, the perturbed value $\tilde{\theta} = \hat{\theta}_n+n^{-1/2}Z$ lies within a small neighborhood of $\theta_0$, and we can write
\[
n^{-1 / 2} U_n(\tilde{\theta})-n^{-1 / 2} U_n(\hat{\theta}_n)=A n^{1 / 2}(\tilde{\theta}-\hat{\theta}_n)+\mathrm{o}_p(1).
\]
Since $U_n(\hat{\theta}_n) = 0$, it follows that
\begin{align}
n^{-1 / 2} U_n(\tilde{\theta})=A Z+\mathrm{o}_p(1).\label{resample}
\end{align}
Consequently, we can numerically estimate each row of the slope matrix $A$ using least squares regression.  Specifically, the $j$-th row of matrix $A$ is estimated by regressing the $j$-th component of $n^{-1 / 2} U_n(\hat{\theta}_n+n^{-1/2}Z)$ on $Z$, using a large number of independent realizations of $Z$. Aggregating these row-wise estimators, we obtain the full matrix estimator $\hat{A}$. Finally, we estimate the asymptotic covariance matrix $\Sigma$ by plugging $\hat{A}$ and $\hat{B}$ into the sandwich formula from Theorem \ref{thm3}, i.e., $\hat{\Sigma} = \hat{A}^{-1}\hat{B}(\hat{A}^T)^{-1}$.

\section{Simulation Studies}\label{sec:5simu}
In this section, we evaluate the finite sample performance of the proposed SMPLE method under both non-homogeneous Poisson process (NHPP) and non-Poisson process models. We begin by considering the NHPP setting with various intensity functions and comparing the estimation accuracy and computational complexity of SMPLE with existing methods. We then study the non-Poisson process, including Gamma frailty models with Cox-type and AFT-type intensity functions, and compare the estimation accuracy of the SMPLE method with the ``reda" package developed by \cite{reda-package}. For ease of reference in the following discussion, we use ODE-Cox, ODE-AM, ODE-LT, and ODE-Flex to denote the SMPLE estimators under the Cox-type model, the AFT-type model, the transformation model with a specified transformation function, and the general linear transformation model with unspecified functional parameters $\alpha(t)$ and $q(t)$, respectively.

\subsection{The Linear Transformation Models}
We first generate the event times from NHPPs with the intensity function satisfying
\[
\mu_x'(t) = q(\mu_x(t))\exp(\beta_1 x_1 + \beta_2 x_2 + \beta_3 x_3)\alpha(t),
\]
where $\beta_1 = \beta_2 = \beta_3 = 1$. We consider four different settings for the functions $\alpha(\cdot)$ and $q(\cdot)$: 1) Cox-type model: $q(t) = 1$ (constant) and  $\alpha(t) = t^2+1$ (monotonically increasing). 2) AFT-type model: $q(t) = \frac{2}{t+1}$ (monotonically decreasing) and  $\alpha(t) = 1$ (constant). 3) Linear transformation model: $q(t) = \frac{1}{t/2+1}$ and $\alpha(t) = \frac{0.2}{1+t}$, both monotonically decreasing. This corresponds to the transformation model in \cite{zeng2007maximum} with a specified Box-Cox transformation and the transformation parameter $\rho=0.5$. 4) General linear transformation model: $q(t) = \frac{2}{t+1}$ (monotonically decreasing) and $\alpha(t) = t+1$ (monotonically increasing).  For settings 1), 2), and 4), the covariates $X_i$ are independently generated from a normal distribution with mean 0 and standard deviation 0.5, truncated at $\pm 4$. For setting 3), $X_1$ and $X_2$ are independently drawn from $N(0,1)$ and truncated at $\pm 1$, while $X_3$ is generated independently from a Bernoulli($0.5$) distribution.

In Setting 1, the censoring times are independently generated from a uniform distribution $U(0, 2)$, resulting in approximately three events per individual on average. We fit $\log \alpha(t)$ using quartic B-splines with $\lceil N^{\frac{1}{7}}\rceil$ interior knots, where $N$ is the total number of observed event times. The interior knots are equally spaced over the interval from $0$ to the maximum observed time. We compare the proposed sieve estimator with the ``cox.LWYY" method by \cite{lin2001transformation}, implemented in the R package ``reReg".
In Setting 2, the censoring times are independently generated from $U(1, 3)$. We fit $\log q(t)$ using cubic B-splines with $\lceil N^{\frac{1}{5}}\rceil$ equally spaced interior knots. The sieve estimator is compared against the ``am.GL" method in the ``reReg" package, which corresponds to the rank-based estimator proposed by \cite{ghosh2003semiparametric}.
In Setting 3, the censoring times are independently generated from $\min\{U(2, 6), 4\}$. We fit $\log \alpha(t)$ using cubic B-splines with $\lceil N^{\frac{1}{5}}\rceil$ interior knots placed at the quantiles of distinct event times. For comparison, we also fit the nonparametric maximum likelihood estimator (NPMLE) developed by \cite{zeng2007maximum}, using the specified Box-Cox transformation function.
In Setting 4, the censoring times are independently generated from $U(1, 3)$. Both $\log \alpha(t)$ and $\log q(t)$ are fitted using cubic B-splines with $\lceil N^{\frac{1}{5}}\rceil$ interior knots located at the quantiles of distinct event times. We impose identifiable constraints  $\beta_1 = 1$ and $\alpha(2) = 1$. 
Additionally, We apply the ODE-Flex method to all four settings, treating both $\alpha(\cdot)$ and $q(\cdot)$ as unspecified. For identifiability, we set $\beta_1=1$ and $\alpha(t_0)=1$, where $t_0$ is the median of all observed event times.

We summarize the results for the ODE-Cox, ODE-AM, and ODE-LT estimators 
in Table \ref{cox_aft_table}, and for the ODE-Flex estimator in Table \ref{ode_flex_table}, based on 1,000 simulation replications with a sample size of $n=1000$. Note that due to the high computational cost, the ``am.GL'' method was evaluated using only 200 replications. Table \ref{cox_aft_table} shows that the proposed sieve estimator performs comparably to existing methods across all specific model settings. In particular, under the Cox-type setting, the ODE-Cox estimator yields nearly identical means and standard errors to the ``cox.LWYY'' estimator; under the AFT-type setting, the ODE-AM estimator exhibits slightly higher bias but similar standard errors compared to the ``am.GL'' estimator; and under the transformation setting, the ODE-LT estimator demonstrates smaller bias and comparable standard errors relative to the NPMLE method.  These results suggest that the proposed SMPLE method can serve as a unified estimation approach across a variety of recurrent event models. For these comparisons, the covariance matrix of the parameter estimates is obtained by inverting the empirical information matrix of the estimated regression and spline coefficients. This is justified under the NHPP setting, where the asymptotic covariance of the proposed estimator equals the Fisher information matrix, as discussed earlier. Results for larger sample sizes  ($n=2000$, $4000$, and $8000$) exhibit similar patterns and are therefor omitted for brevity. 

\begin{figure}[ht]
     \centering
     \captionsetup{font=scriptsize}
     \begin{subfigure}[]{0.4\textwidth}
         \centering
         \caption*{Setting (2)}
         \includegraphics[width=\textwidth]{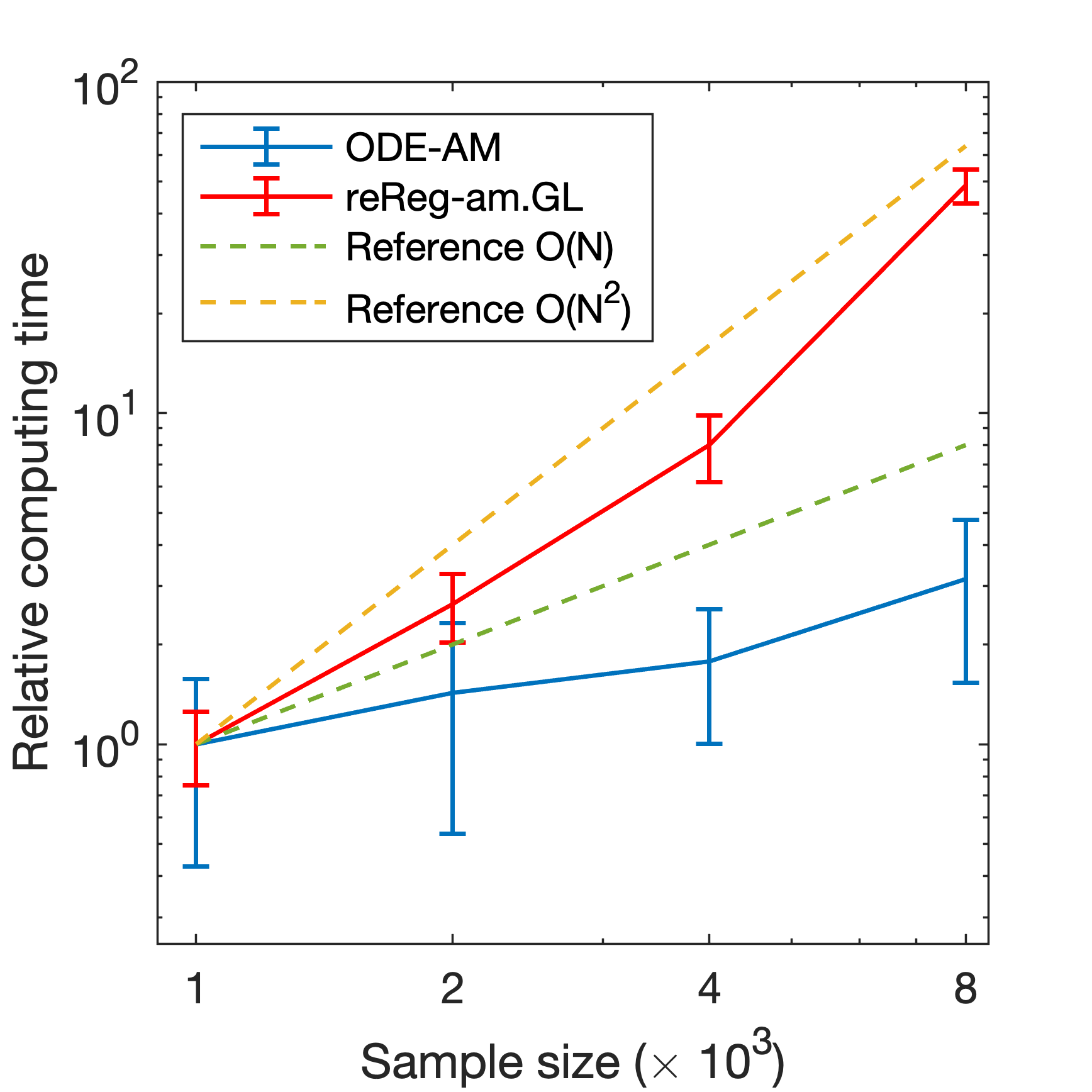}
     \end{subfigure}
     \begin{subfigure}[]{0.4\textwidth}
         \centering
         \caption*{Setting (3)}
         \includegraphics[width=\textwidth]{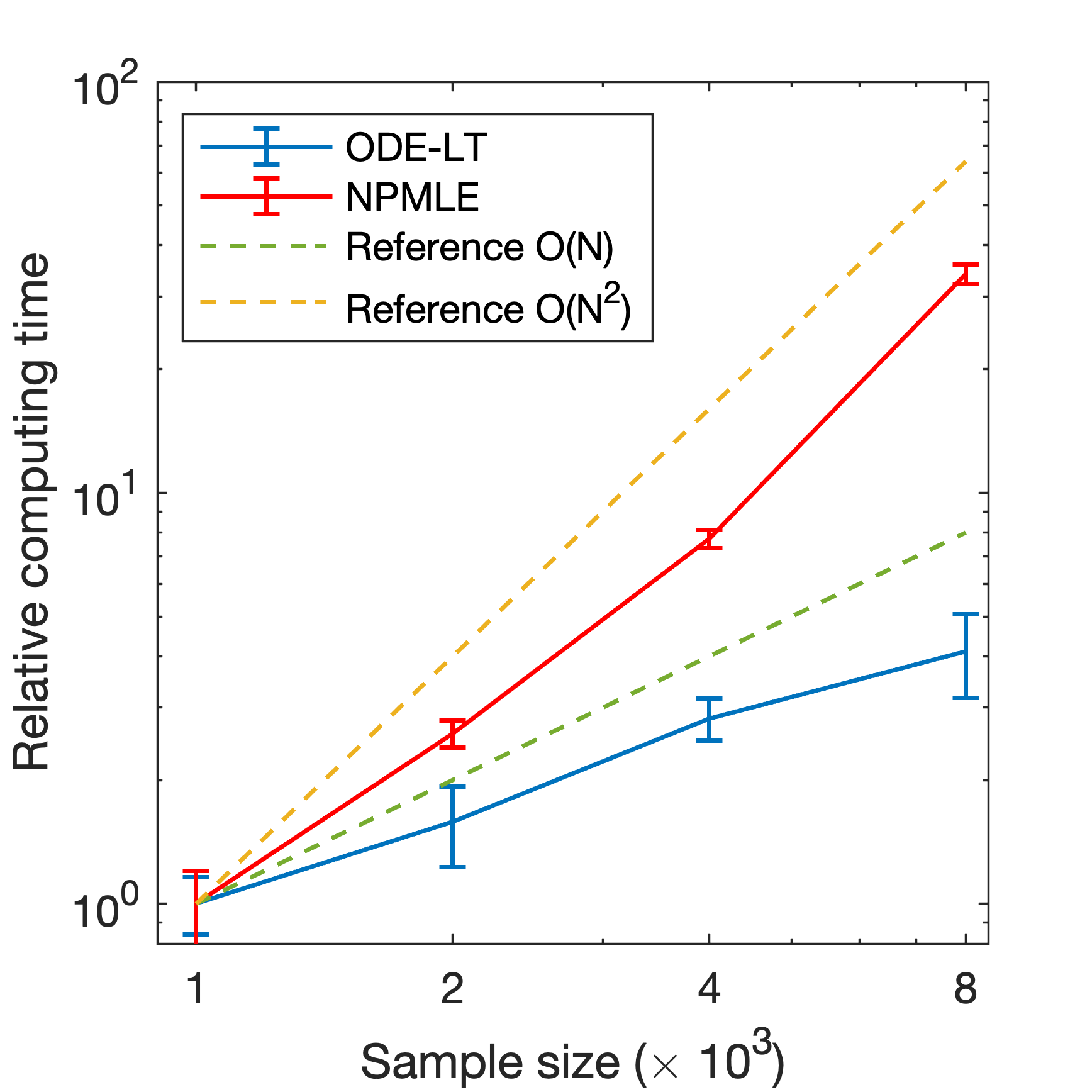}
     \end{subfigure}
          \captionsetup{font=normal}
             \caption{Left: the log-log plot of relative computation time with respect to sample size $n$ under the AFT-type model and the transformation model with specified $q(\cdot)$ from left to right, respectively.}
        \label{time_compare}
\end{figure}

\begin{figure}[ht]
     \centering
     \captionsetup{font=scriptsize}
     \begin{subfigure}[]{0.32\textwidth}
         \centering
         \caption*{Setting (1)}
         \includegraphics[width=\textwidth]{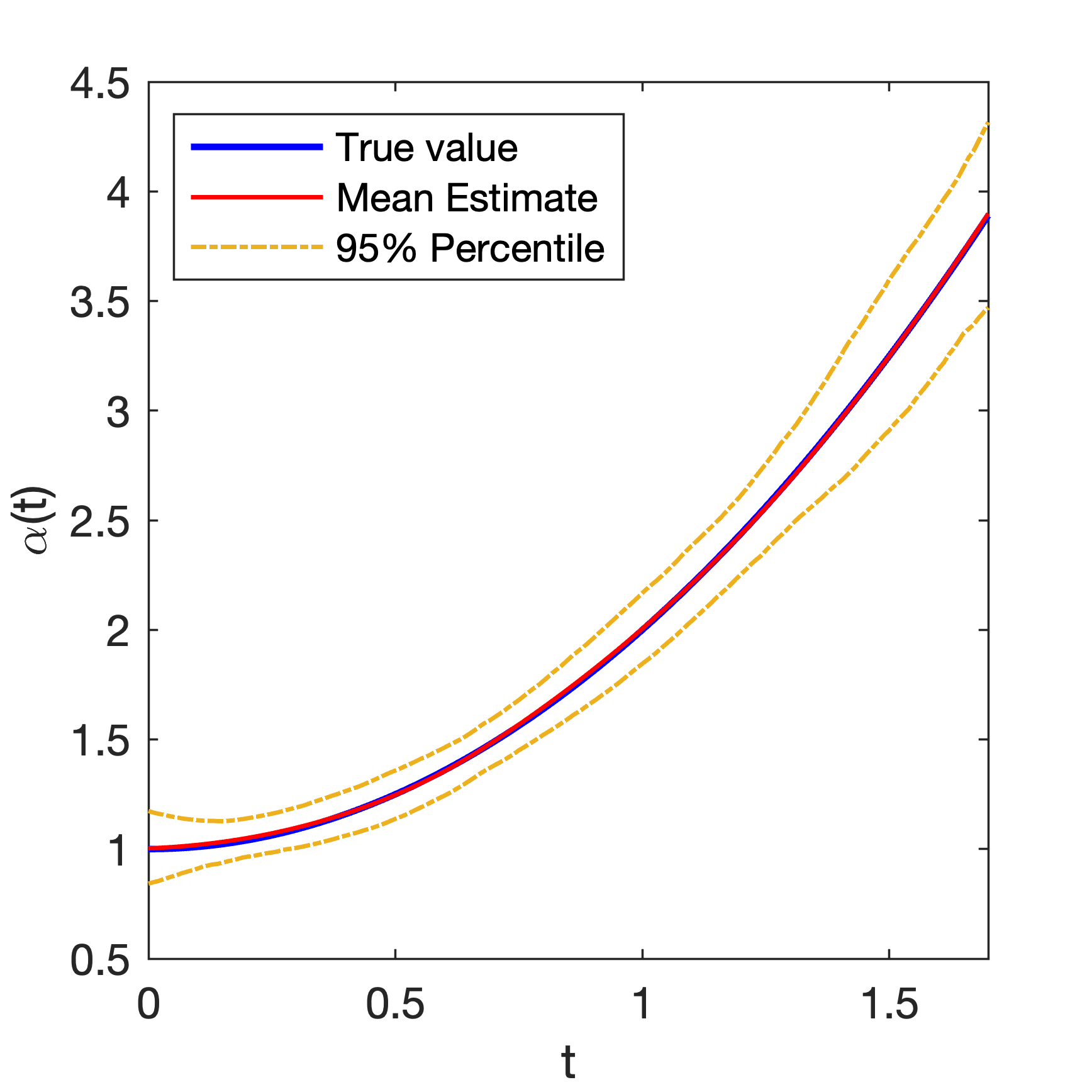}
     \end{subfigure}
     \begin{subfigure}[]{0.32\textwidth}
         \centering
         \caption*{Setting (2)}
         \includegraphics[width=\textwidth]{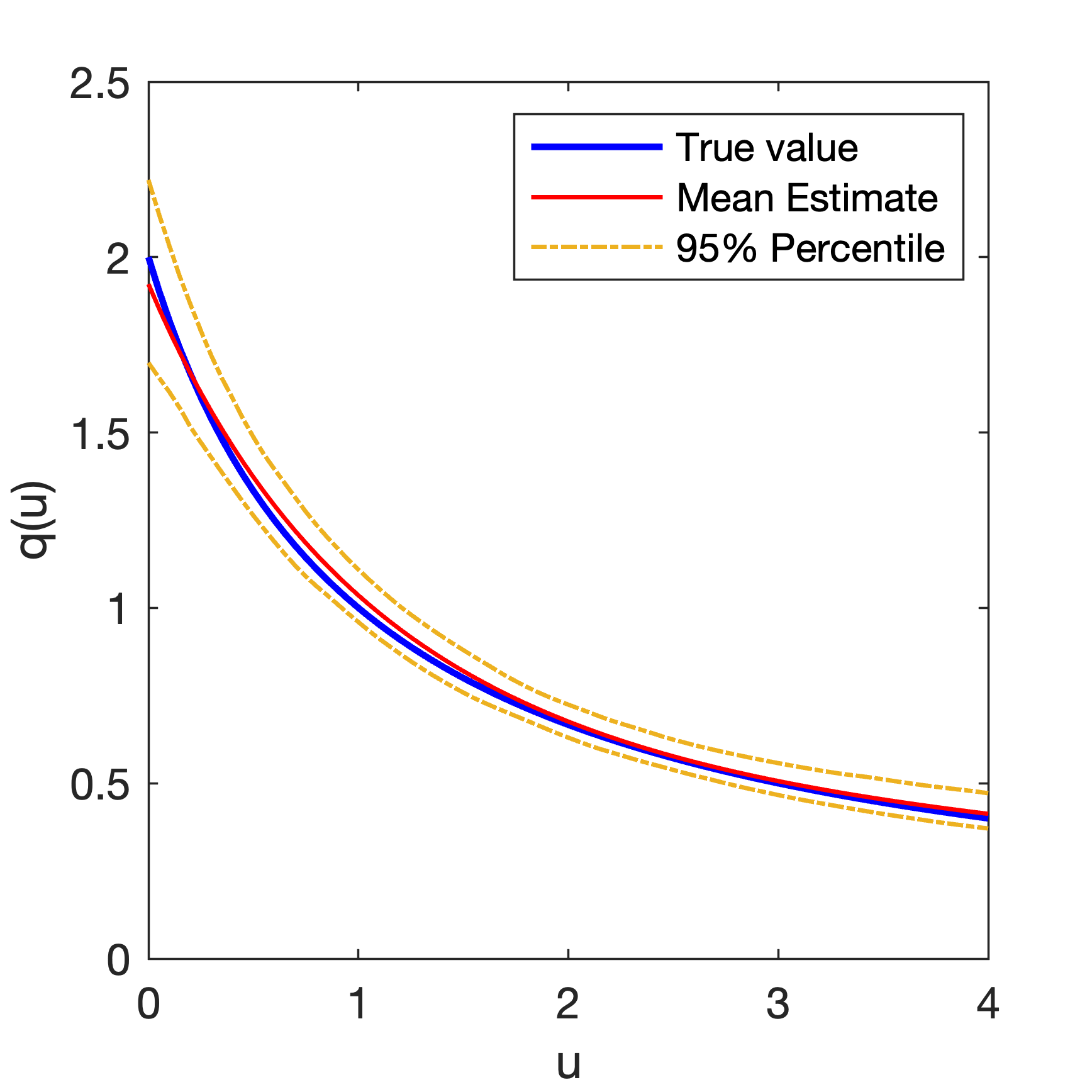}
     \end{subfigure}
          \begin{subfigure}[]{0.32\textwidth}
         \centering
         \caption*{Setting (3)}
         \includegraphics[width=\textwidth]{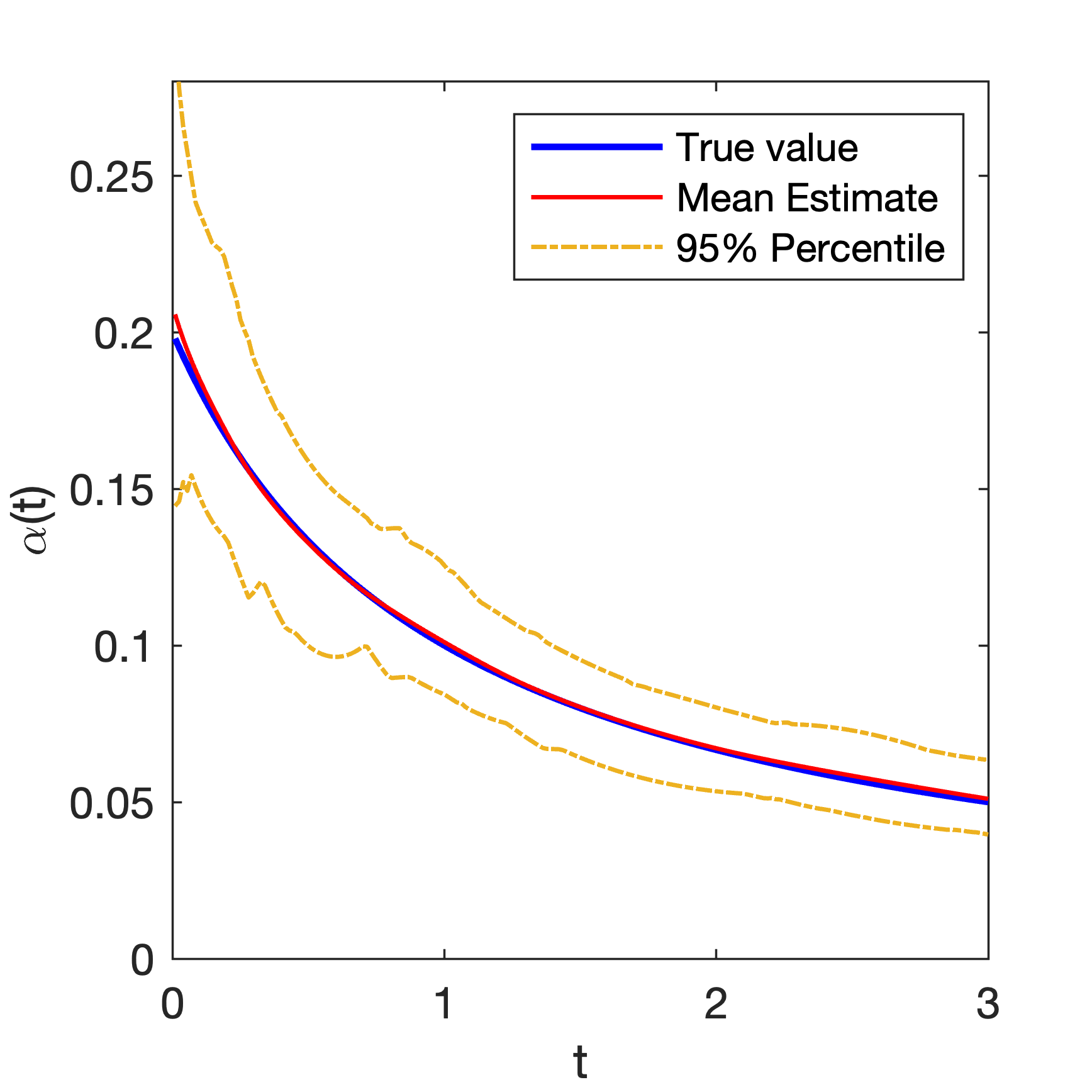}
     \end{subfigure}
     \captionsetup{font=normal}
        \caption{True $\alpha_0(t)$, mean of spline estimators $\hat{\alpha}(t)$, and the $95\%$ pointwise confidence intervals with sample size $n = 1000$.}
        \label{cox_summary}
\end{figure}

Despite the similar statistical performance across various settings, the proposed sieve estimator exhibits significantly better  computational efficiency  compared to 
the ``am.GL" method under the AFT-type setting and the NPMLE under the linear transformation setting. As shown in Figure \ref{time_compare}, the relative computation time of the proposed SMPLE method consistently grows at a sublinear rate as sample size increases. In contrast, the computation time for both the ``am.GL'' method and the NPMLE approach approximate a quadratic growth rate, making them substantially slower for large datasets. 
Figure \ref{cox_summary} presents the spline estimators $\hat{\alpha}(t)$ for the Cox-type model, $\hat{q}(u)$ for the AFT-type model, and $\hat{\alpha}(t)$ for the linear transformation model with a specified  functional parameter $q(u)$. The results indicate that the means of $\hat{\alpha}(t)$ and $\hat{q}(u)$ across replicates accurately approximate the underlying true functions, and the $95\%$ pointwise confidence bands cover the true functions well across different settings.
\begin{table}[ht]
\caption{Simulation results under correctly specified ODE-Cox with $q(\cdot) \equiv 1$, ODE-AM with $\alpha(\cdot) \equiv 1$, ODE-LT with $q(t) = \frac{1}{t/2+1}$.}
\begin{center}
\begin{tabular}{ cc|cccc|cccc }
\toprule[1.2pt]
 \multicolumn{1}{c}{Setting}& \multicolumn{1}{c|}{Method} &\multicolumn{4}{c|}{$\beta_2 = 1$}&\multicolumn{4}{c}{$\beta_3 = 1$}\\
&&Bias&SE&ESE&CP&Bias&SE&ESE&CP\\
\hline
\multirow{2}{*}{1)} &ODE-Cox&   -0.002  &  0.039 &   0.043  &  0.958&-0.002  &  0.041  &  0.043  &  0.948\\
&reReg-Cox& -0.002& 0.039& 0.040& 0.956&-0.002& 0.041& 0.041& 0.941\\
\hline
\multirow{2}{*}{2)} &ODE-AM&      -0.021 &0.064 &   0.063  &  0.936&    -0.022  &  0.063  &  0.063  &  0.943 \\
&reReg-AFT& 0.005&0.062&-&-&-0.000&0.064&-&-\\
\hline
\multirow{2}{*}{3)} &ODE-LT&     0.000 &0.078&    0.072 &   0.960&0.003   & 0.088 &   0.100&    0.980\\
&NPMLE&0.017 &   0.074  &  0.078 &   0.960&0.024   & 0.101 &   0.109 &   0.962\\
\toprule[1.2pt]
\end{tabular}
\label{cox_aft_table}
\captionsetup{font=scriptsize}
\caption*{Bias is the difference between the mean of the estimated coefficients and the true values.
SE is the standard deviation of the estimates. ESE is the mean of the standard error by inverting the estimated information matrix of the regression and the spline coefficients. CP is the proportion of the estimated $95\%$ confidence intervals that cover the true parameters.}
\end{center}
\end{table}

\begin{figure}[ht]
     \centering
     \captionsetup{font=scriptsize}
      \begin{subfigure}[]{0.22\textwidth}
         \centering
         \caption*{Setting (1)}
         \includegraphics[width=\textwidth]{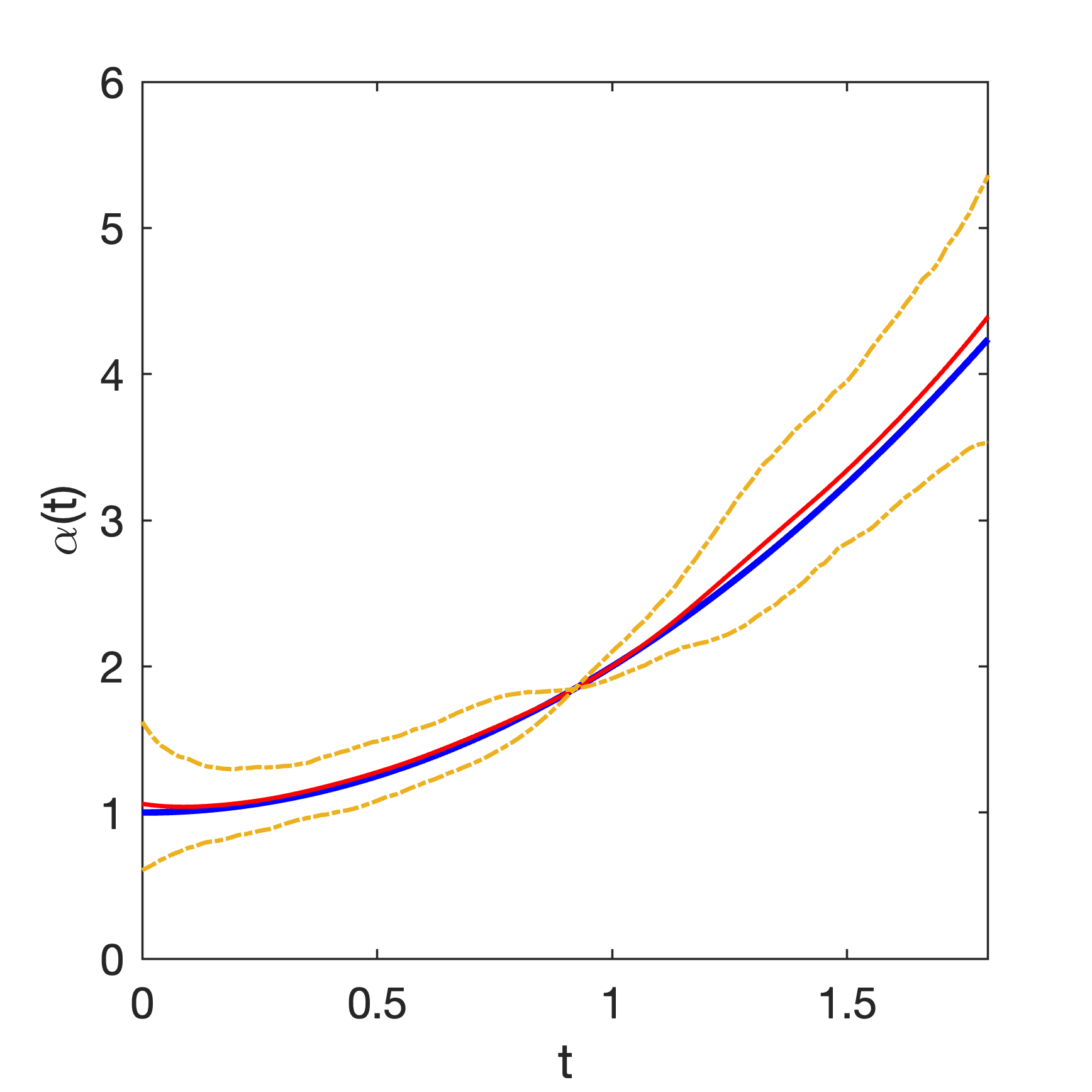}
     \end{subfigure}
     \hfill
     \begin{subfigure}[]{0.22\textwidth}
         \centering
         \caption*{Setting (2)}
         \includegraphics[width=\textwidth]{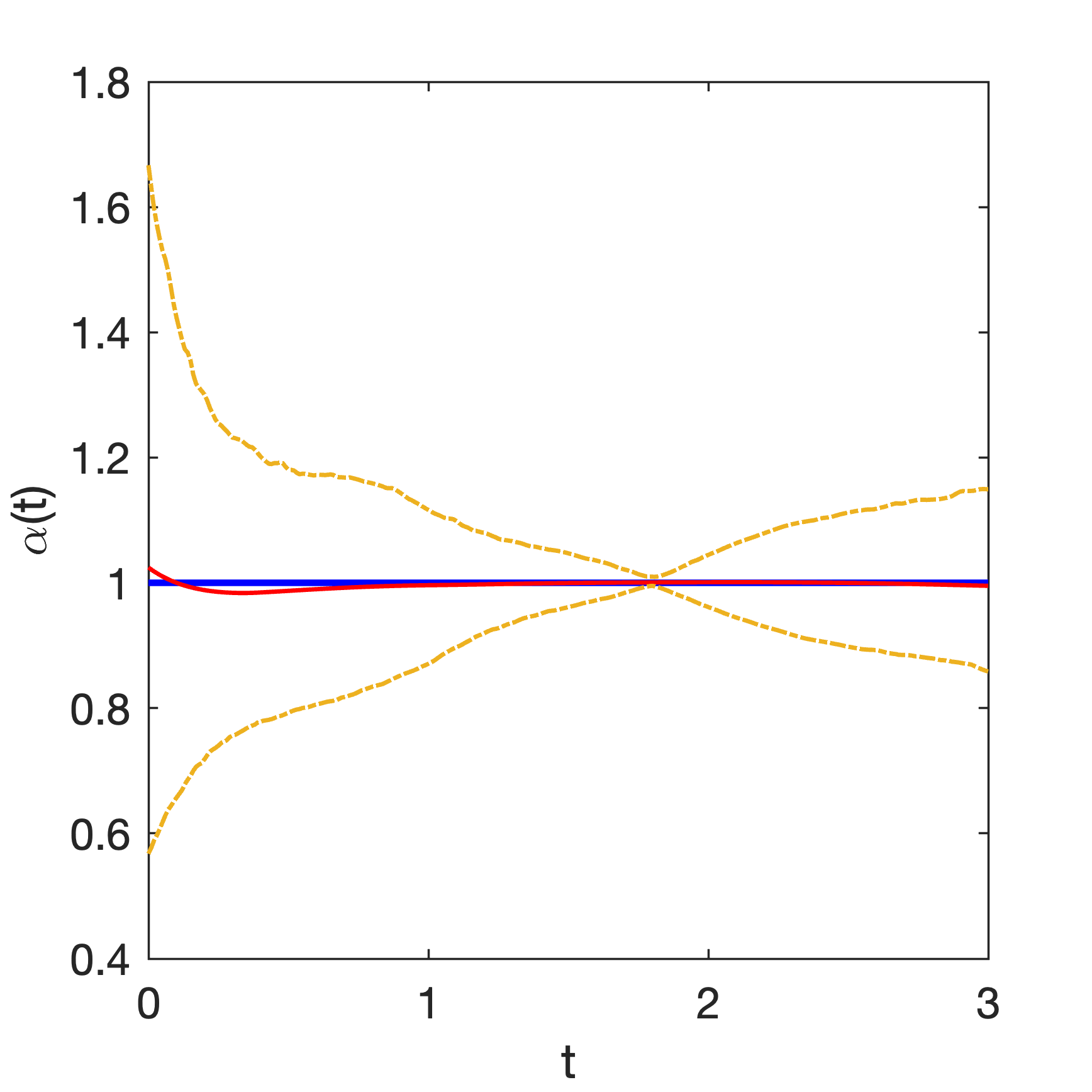}
     \end{subfigure}
     \hfill
     \begin{subfigure}[]{0.22\textwidth}
         \centering
         \caption*{Setting (3)}
         \includegraphics[width=\textwidth]{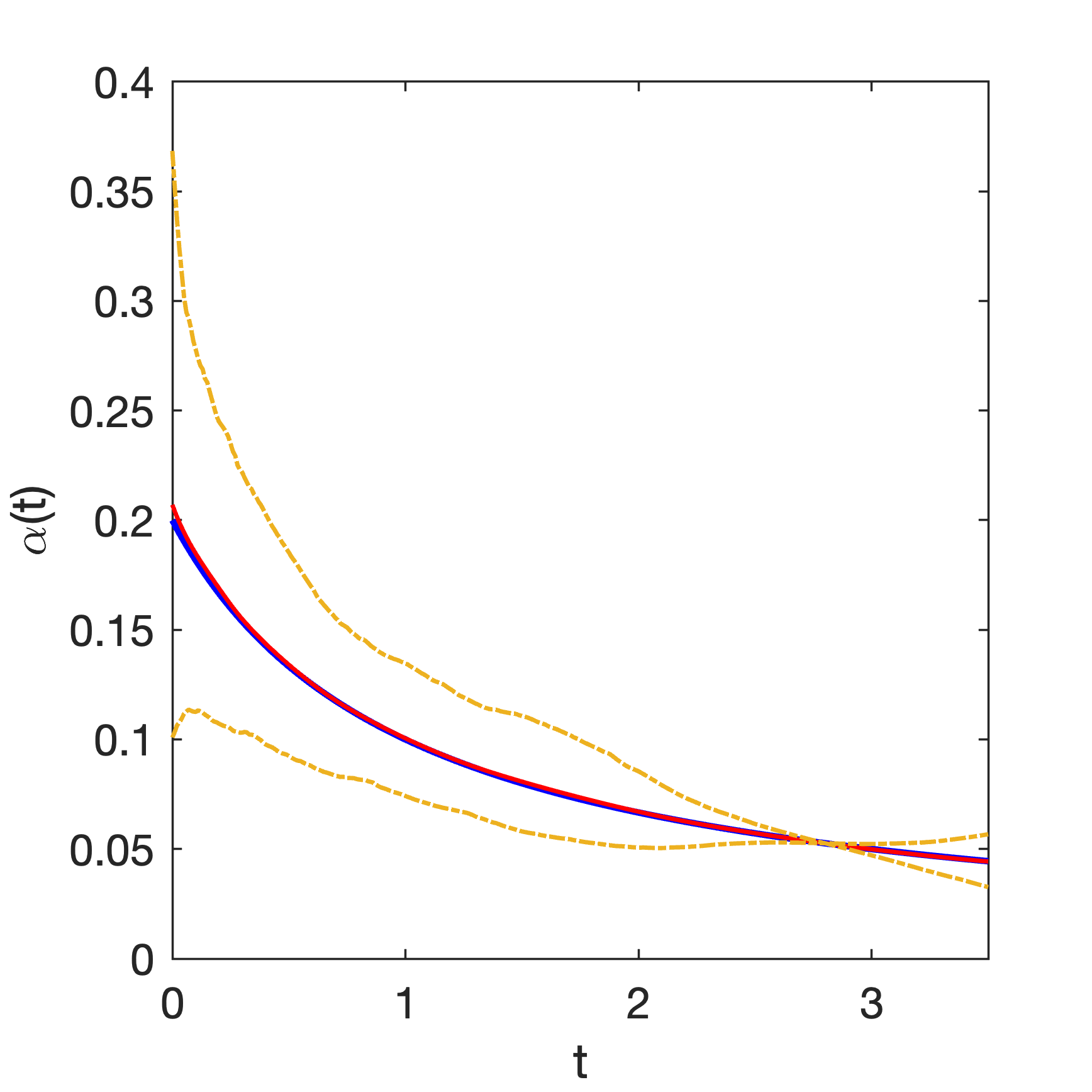}
     \end{subfigure}
     \hfill
     \begin{subfigure}[]{0.22\textwidth}
         \centering
         \caption*{Setting (4)}
         \includegraphics[width=\textwidth]{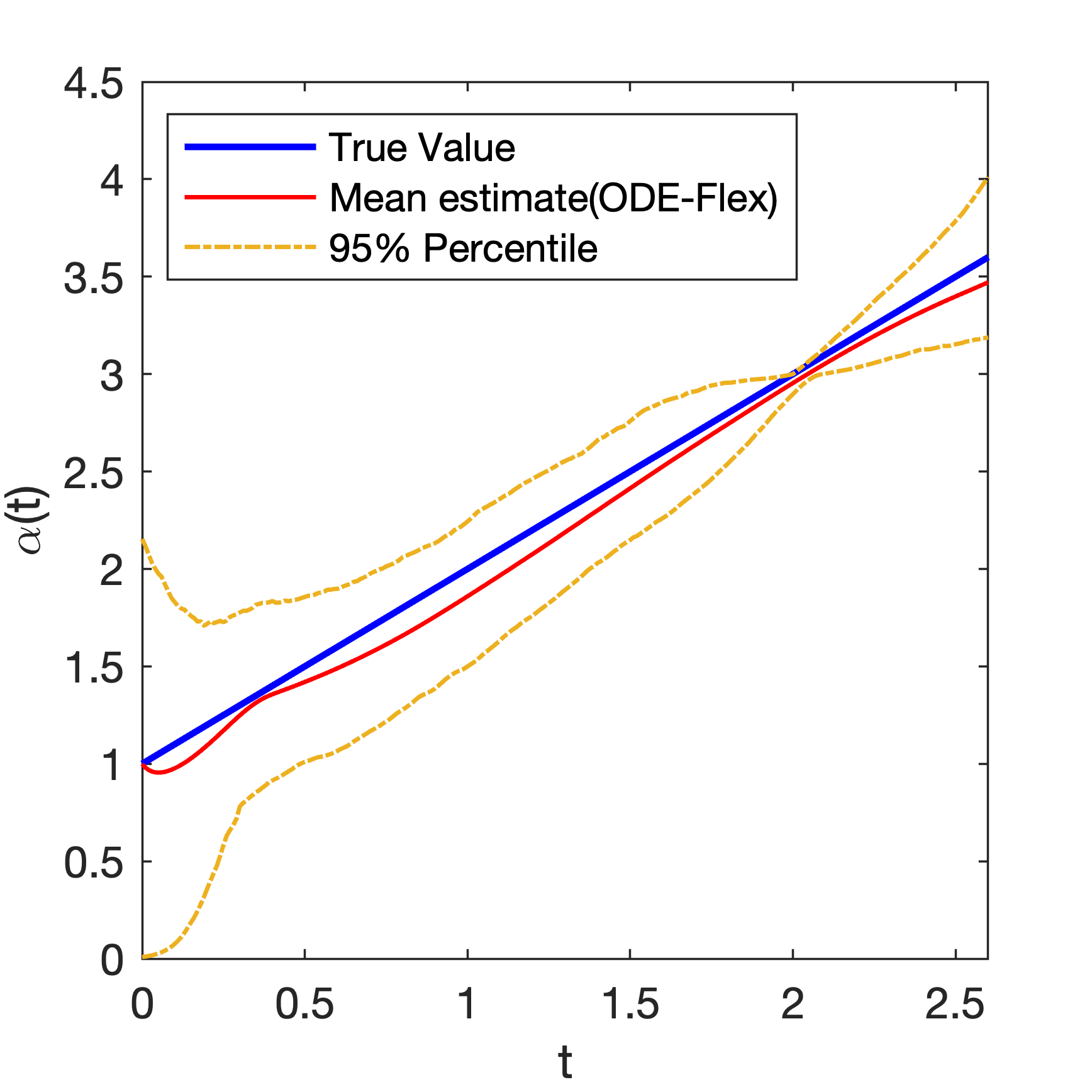}
     \end{subfigure}
          \begin{subfigure}[]{0.22\textwidth}
         \centering
         \includegraphics[width=\textwidth]{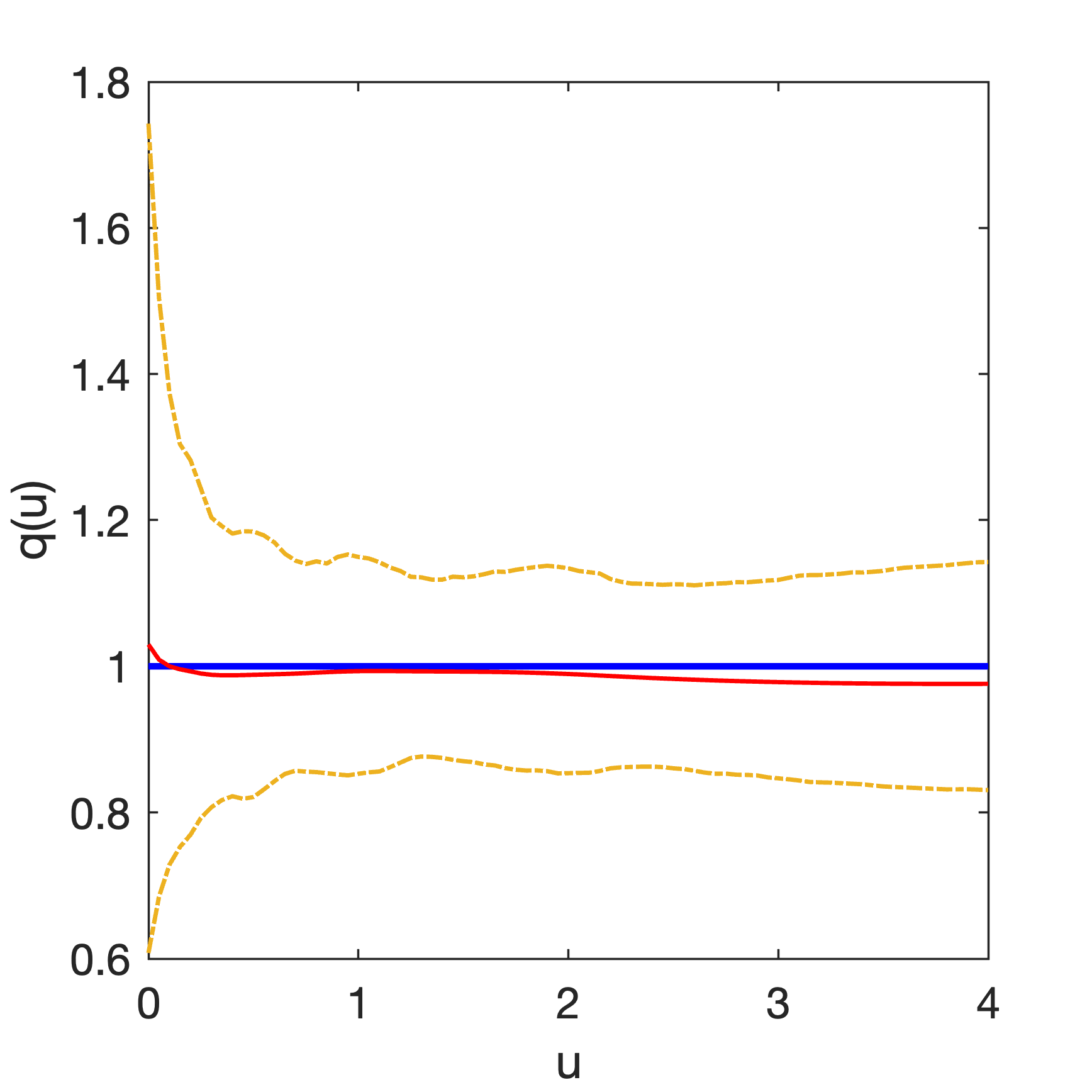}
     \end{subfigure}
     \hfill
        \label{aft_plot}
     \begin{subfigure}[]{0.22\textwidth}
         \centering
         \includegraphics[width=\textwidth]{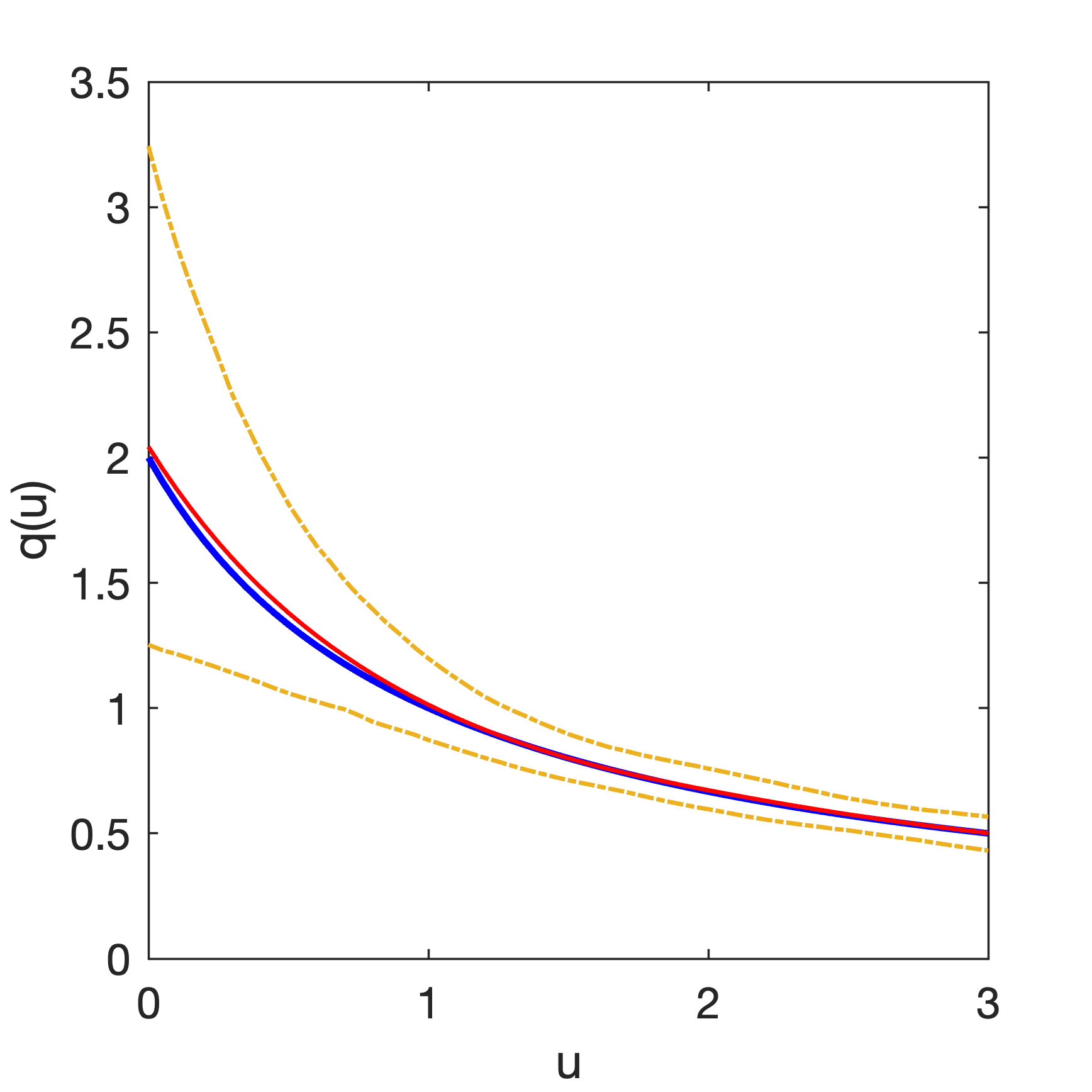}
     \end{subfigure}
     \hfill
     \begin{subfigure}[]{0.22\textwidth}
         \centering
         \includegraphics[width=\textwidth]{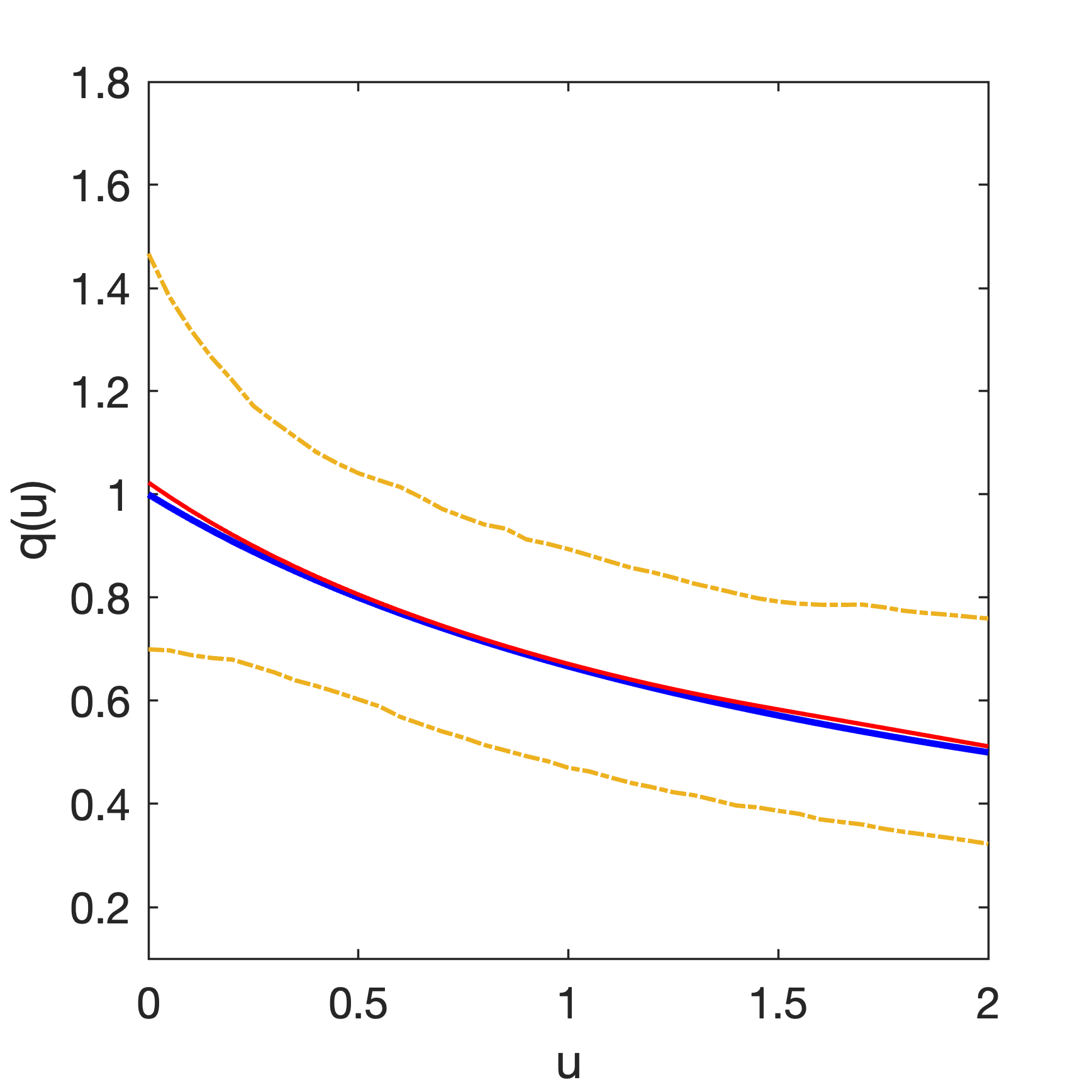}
     \end{subfigure}
     \hfill
     \begin{subfigure}[]{0.22\textwidth}
         \centering
         \includegraphics[width=\textwidth]{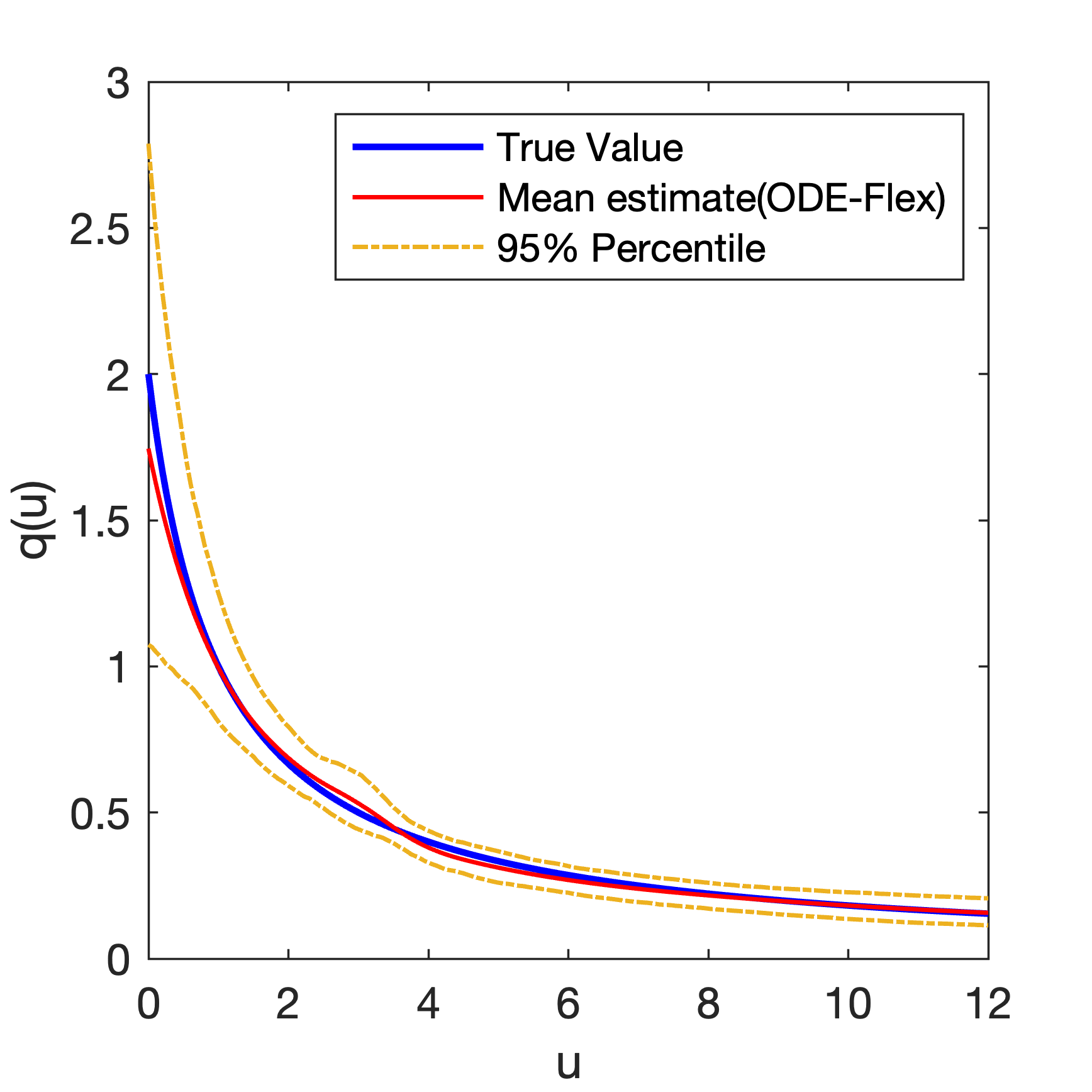}
     \end{subfigure}
              \captionsetup{font=normal}
        \caption{The solid red curve represents the true $\alpha(\cdot)$ (upper row) and true $q(\cdot)$ (lower row). The solid blue curve represents the average spline estimator $\hat{\alpha}(\cdot)$ and $\hat{q}(\cdot)$ over 1000 replications. The dashed yellow curves represent the $95\%$ confidence bands over the 1000 replications. From left to right, the columns represent model settings (1)-(4) respectively.}
        \label{ode_flex_plot}
\end{figure}

In addition to fitting recurrent event models with specified functional parameters, we also applied the ODE-Flex estimator with unspecified $\alpha(\cdot)$ and $q(\cdot)$ under settings (1)-(4). As shown in Table \ref{ode_flex_table}, the ODE-Flex estimator accurately recovers the true regression parameters, exhibiting negligible bias, and the estimated $95\%$ confidence intervals achieve reasonable coverage probabilities. Furthermore, when the model is correctly specified or when functional parameters are partially known, the corresponding specified estimators generally achieve smaller standard errors compared to the more flexible ODE-Flex estimator. For example, the SEs reported in Table \ref{cox_aft_table} are consistently lower than those in Table \ref{ode_flex_table} under the same model settings. Figure \ref{ode_flex_plot} illustrates that the averaged spline estimators $\hat{\alpha}(\cdot)$ and $\hat{q}(\cdot)$ under the general linear transformation model closely approximate the true functions, with the $95\%$ pointwise confidence bands covering the true curves. 
Notably, all curves for $\hat{\alpha}(\cdot)$ in Figure \ref{ode_flex_plot} intersect at a single ``node'' point due to the identifiability constraint $\alpha(t_0)=1$. 
As a result, $\hat{\alpha}(t_0)$ takes the same value across all replications. Moreover, since the constraint anchors the function at $t_0$, the variance of $\hat{\alpha}(t)$ is reduced in its vicinity, making the estimator more stable near this point. For visual clarity and consistency, we rescale $\hat{\alpha}(t)$ such that $\hat{\alpha}(t_0) = \alpha_0(t_0)$ and present the rescaled $\hat{\alpha}(t)$ and $\hat{q}(u)$ in Figure~\ref{ode_flex_plot}.
\begin{table}[ht]
\begin{center}
\caption{Estimated regression coefficients with ODE-Flex under the general linear transformation models with $\alpha(\cdot)$ and $q(\cdot)$ unspecified. Bias, SE, ESE, and CP have the same meanings as in Table \ref{cox_aft_table}.}
\begin{tabular}{ c|cccc|cccc }
\toprule[1.2pt]
 \multicolumn{1}{c|}{Setting} &\multicolumn{4}{c|}{$\beta_2 = 1$}&\multicolumn{4}{c}{$\beta_3 = 1$}\\
&Bias&SE&ESE&CP&Bias&SE&ESE&CP\\
\hline
\multirow{1}{*}{1)} &     0.005   & 0.059 &   0.061  &  0.952&0.006  &  0.059 &   0.060   & 0.961\\
\hline
\multirow{1}{*}{2)}  &      -0.002   & 0.083   & 0.079  &  0.933&-0.003 &   0.084   & 0.079   & 0.936\\
\hline
 \multirow{1}{*}{3)} & 0.001 &   0.097  &  0.092  &  0.934&
    0.006  &  0.119 &   0.116  &  0.949\\
    \hline
 \multirow{1}{*}{4)} & -0.007 &   0.071  &  0.066 &   0.939&
   -0.006  &  0.071&    0.066   & 0.926\\
\toprule[1.2pt]
\end{tabular}
\label{ode_flex_table}
\end{center}
\end{table}

\subsection{Gamma Frailty Models}
In this section, we evaluate the performance of the SMPLE method under non-Poisson processes by considering Gamma frailty models. Specifically, suppose the intensity function of the recurrent event process conditional on covariates $X$ is given by $\rho_x(t) = \xi \mu_x'(t)$, where the random effect $\xi$ follows a Gamma distribution with mean 1 and variance 0.5, and the conditional mean function $\mu_x(t)$ satisfies the following ODE:
\[
\mu_x'(t) = q(\mu_x(t))\exp(\beta_1 x_1 + \beta_2 x_2 + \beta_3 x_3)\alpha(t),
\] 
where $\beta_1 = \beta_2 = \beta_3 = 1$. For the simulation study, the covariates
$X$ are generated in the same way as in Setting 1 in the previous subsection. We consider two model configurations for the functional parameters $\alpha(\cdot)$ and $q(\cdot)$. Setting 5): $q(t) = 1$ (constant) and  $\alpha(t) = t^2+1$ (monotonically increasing); Setting 6): $q(t) = \frac{2}{t+1}$ (monotonically decreasing) and $\alpha(t) = 1$ (constant). To estimate the covariance matrix of the coefficient estimators, we  apply Theorem 3 directly for the ODE-Cox estimator. For ODE-AM and ODE-Flex estimators, we employ the resampling method described in Section~\ref{sec:sec4_thms} to compute the covariance matrix numerically. Specifically, for the ODE-AM estimator, we set the number of resampling replications to $B = 100$. For the ODE-Flex estimator, we use $B=1500$ and $B=2000$ for Settings 5) and 6), respectively. These values were selected based on empirical evidence to ensure accurate estimation of confidence intervals. For comparison, we also fit the data under both settings using the ``reda'' package developed by \cite{reda-package}.

\begin{figure}[ht]
     \centering
     \captionsetup{font=scriptsize}
     \begin{subfigure}[]{0.24\textwidth}
         \centering
         \caption*{Setting (5)}
         \includegraphics[width=\textwidth]{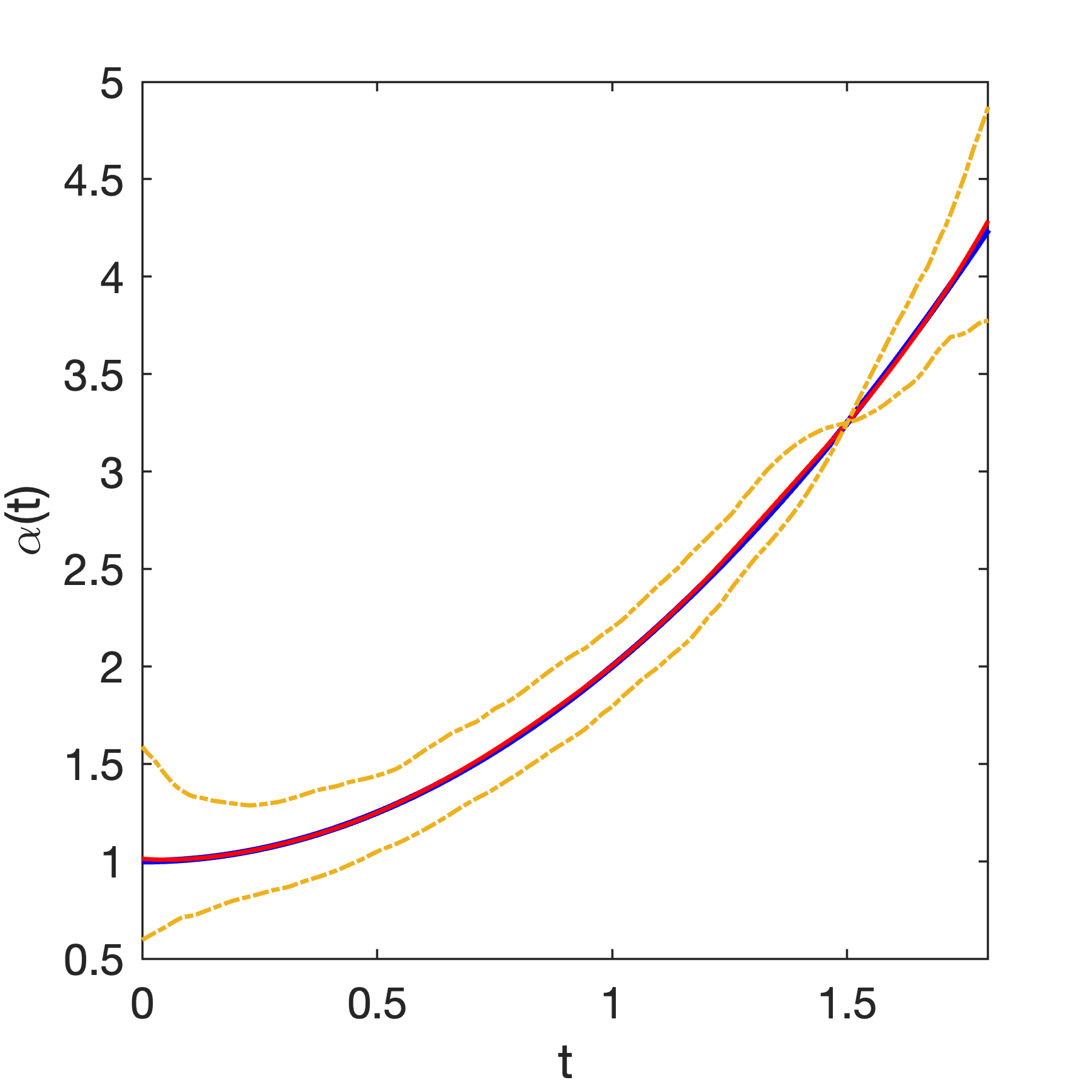}
     \end{subfigure}
     \hfill
     \begin{subfigure}[]{0.24\textwidth}
         \centering
                  \caption*{Setting (5)}
         \includegraphics[width=\textwidth]{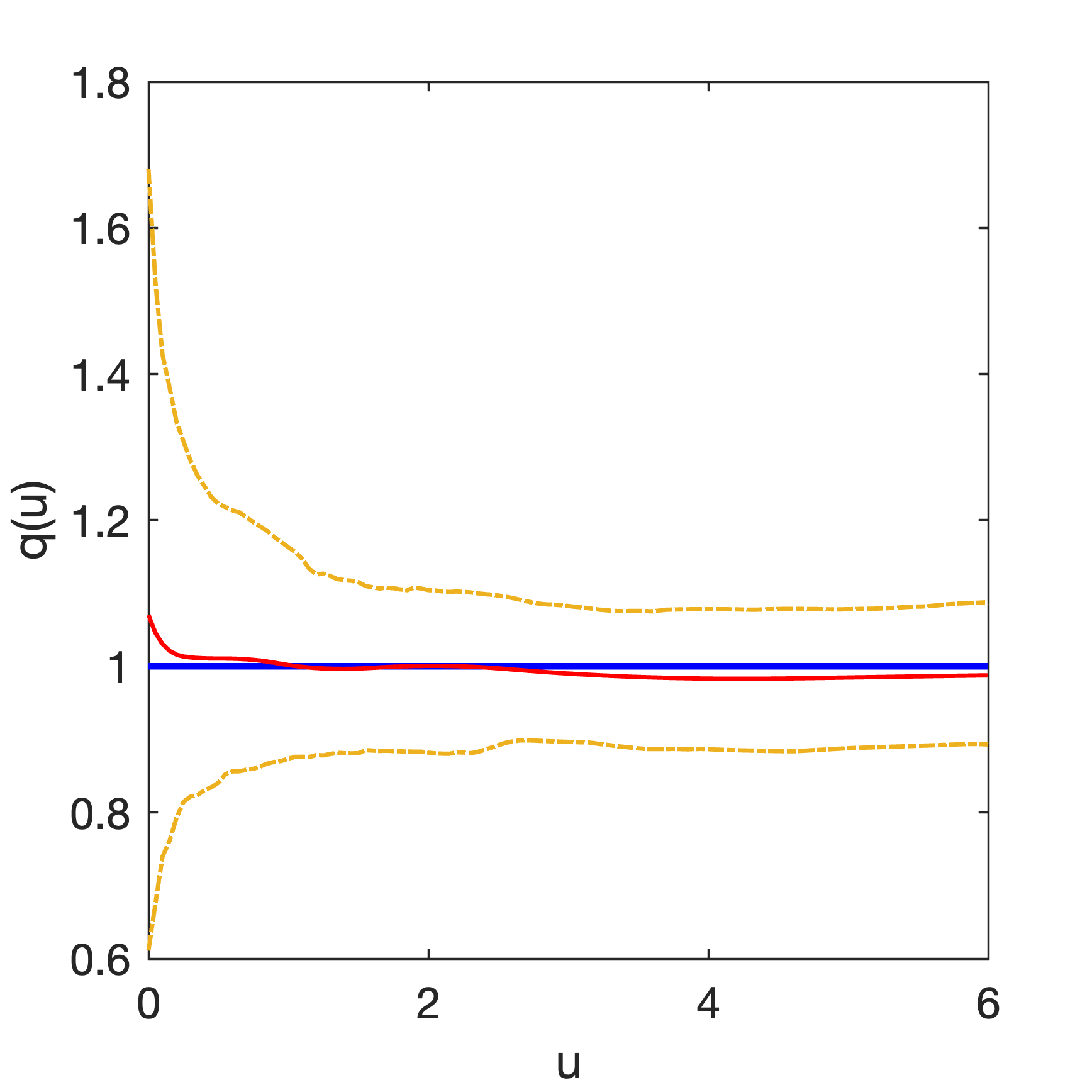}
     \end{subfigure}
     \begin{subfigure}[]{0.24\textwidth}
         \centering
                  \caption*{Setting (6)}
         \includegraphics[width=\textwidth]{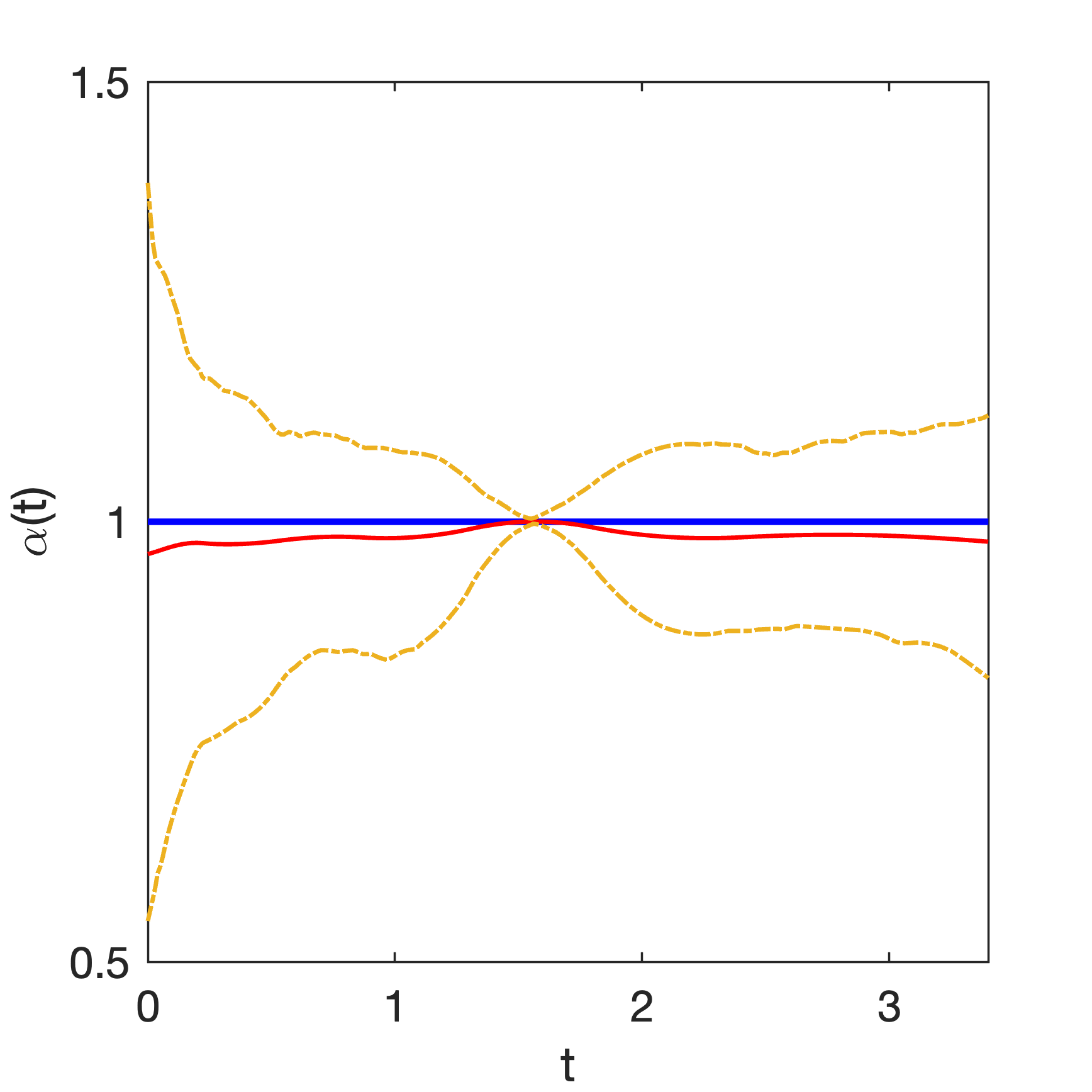}
     \end{subfigure}
     \hfill
     \begin{subfigure}[]{0.24\textwidth}
         \centering
                  \caption*{Setting (6)}
         \includegraphics[width=\textwidth]{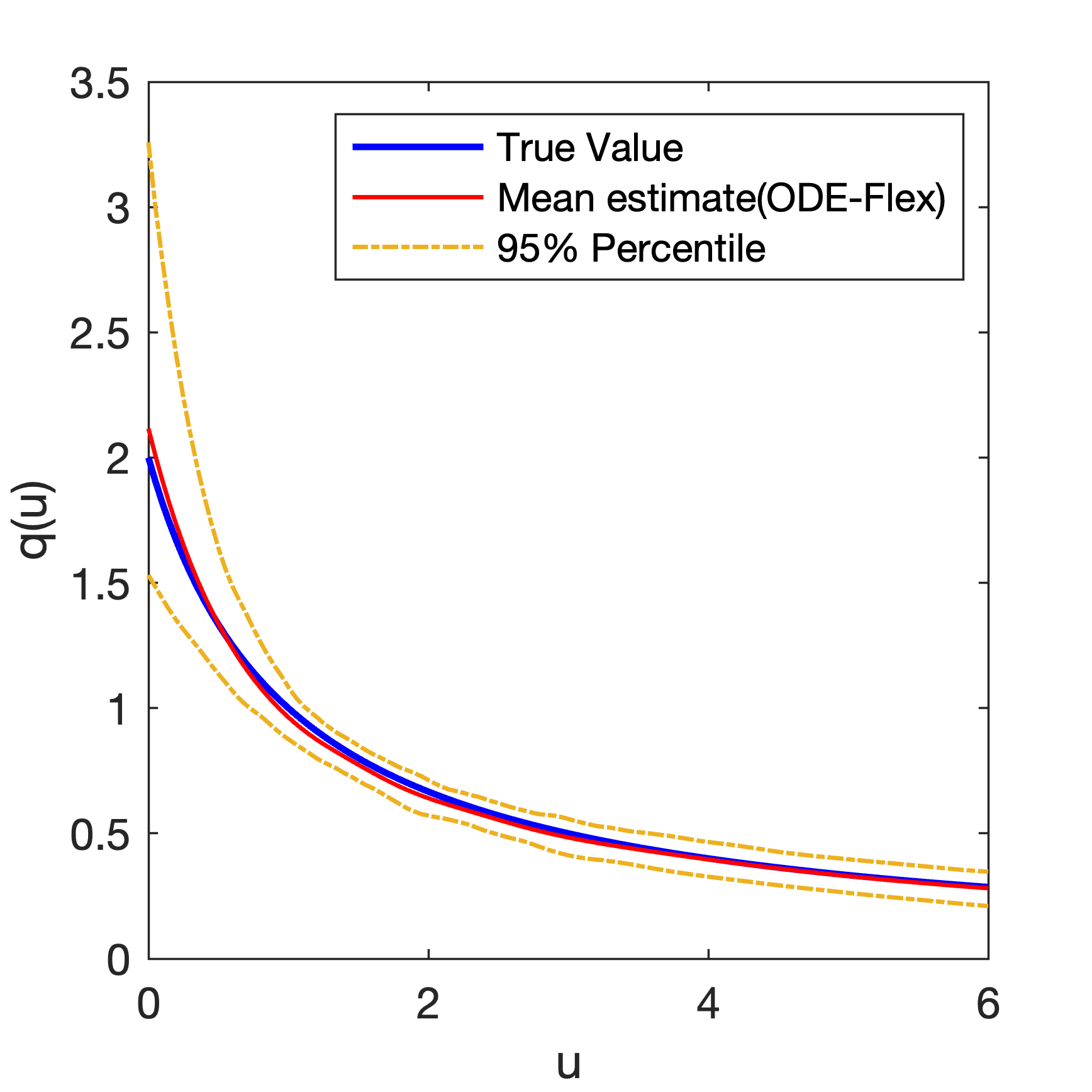}
     \end{subfigure}
          \captionsetup{font=normal}
             \caption{The curves have the same meanings as those in Figure \ref{ode_flex_plot}. The left two columns correspond to setting (5), and the right two columns correspond to setting (6).}
        \label{ltm_gamma_2}
\end{figure}

Table \ref{gamma_table} summarizes the estimation results for sample size $n=2000$ and $n=4000$, based on 1000 replications. Under the Gamma frailty model with a Cox-type intensity function, the proposed sieve estimator performs comparably to the ``reda'' package.  Notably, the standard error (SE) 
of the ``reda'' estimator is the smallest among all the methods, as it assumes the true recurrent event process follows a Gamma frailty model with a Cox-type intensity function. In contrast, the proposed ODE-Cox model requires only that the marginal mean function satisfies a Cox-type ODE; it does not assume a Gamma distribution for the random effects or specify the full intensity structure of the recurrent event process. As a result, the ESE of ODE-Cox is slightly higher than that of ``reda''. On the other hand, the ODE-Flex model, which leaves both $\alpha(\cdot)$ and $q(\cdot)$ unspecified, contains the least information about the underlying process and thus exhibits the largest ESE among the methods considered. When the true recurrent event process follows a Gamma frailty model with an AFT-type intensity function, the ``reda'' method fails to recover the parameters accurately and yields biased estimates. In contrast, the proposed ODE-AFT and ODE-Flex estimators remain robust and perform well. As expected, the ESE of ODE-Flex is larger than that of ODE-AM due to the additional uncertainty introduced by the unspecified $\alpha(\cdot)$. Furthermore, as shown in Figure \ref{ltm_gamma_2}, the spline estimators $\hat{\alpha}$ and $\hat{q}$ obtained from ODE-Flex  accurately approximate the true nuisance functions under both settings, and the $95\%$ pointwise confidence bands consistently cover the true functions.
\begin{table}[ht]
\begin{center}
\caption{Estimates of regression coefficients with data generated from Cox-type and AFT-type model with gamma frailty}
\begin{tabular}{ cc|cccc|cccc }
\toprule[1.2pt]
 \multicolumn{2}{c|}{n=2000} &\multicolumn{4}{c|}{$\beta_2 = 1$}&\multicolumn{4}{c}{$\beta_3 = 1$}\\
Setting&Method&Bias & SE& ESE & CP&Bias & SE& ESE & CP\\
\hline
\multirow{5}{*}{} 
\multirow{3}{*}{Cox} 
&ODE-Cox&  0.001& 0.072 &0.066 &0.937  &-0.001 & 0.069& 0.066 &  0.929\\
&ODE-Flex&   0.006  &  0.088  &  0.107&    0.958  &0.004  &  0.088 &   0.107   & 0.958   \\
&Reda &0.002 & 0.053& 0.052& 0.948& -0.000 & 0.053 &0.052 &0.942 \\
\hline
\multirow{3}{*}{AM} 
&ODE-AM&  -0.004   & 0.070 &   0.070 &   0.945 &-0.005  &  0.070 &   0.070 &   0.948\\
&ODE-Flex&  0.005 &   0.085 &   0.110   & 0.968  & 0.002 &   0.089 &   0.108   & 0.960 \\
&Reda & -0.328 &0.038 &0.042 & 0 &-0.329&0.041 &0.042&0 \\
\toprule[1.2pt]
\end{tabular}

\begin{tabular}{ cc|cccc|cccc }
\toprule[1.2pt]
\multicolumn{2}{c|}{n=4000} &\multicolumn{4}{c|}{$\beta_2 = 1$}&\multicolumn{4}{c}{$\beta_3 = 1$}\\
Setting&Method&Bias & SE& ESE & CP&Bias & SE& ESE & CP\\
\hline
\multirow{5}{*}{} 
\multirow{3}{*}{Cox} 
&ODE-Cox&  0.000& 0.052 &0.048 &0.941  &-0.000 & 0.051& 0.048 &  0.942\\
&ODE-Flex& 0.004   & 0.065 &   0.073  &  0.962  &0.007  &  0.064 &   0.074 &   0.955   \\
&Reda &0.001 & 0.038& 0.037& 0.934& -0.001 & 0.036 &0.036 &0.954 \\
\hline
\multirow{3}{*}{AM} 
&ODE-AM&   -0.003   & 0.050 &   0.051  &  0.945 &-0.004   & 0.050  &  0.051 &   0.950\\
&ODE-Flex&      -0.001  &  0.064 &   0.086    &0.963  &0.001  &  0.063  &  0.087  &  0.967 \\
&Reda & -0.331 &0.029 &0.030 & 0 &-0.330&0.028 &0.030&0 \\
\toprule[1.2pt]
\end{tabular}
\label{gamma_table}
\end{center}
\end{table}

\section{Real Data Example}\label{sec:real data6}
In this section, we demonstrate the application of the proposed SMPLE method using the Medical Information Mart for Intensive Care (MIMIC-III) dataset, as described by \cite{johnson2016mimic}. Developed by the MIT Lab for Computational Physiology, MIMIC-III is a publicly accessible, de-identified database containing comprehensive clinical data from Intensive Care Unit (ICU) stays at the Beth Israel Deaconess Medical Center in Boston, Massachusetts, spanning the years 2001 to 2012. Our primary objective is to identify key medical features that are associated with the frequency of patient readmission to the ICU.

The dataset includes clinical records from 49,785 hospital admissions involving 38,597 unique adult patients aged 15 and older. For each patient, we define the event times as the duration (in years) from their initial hospitalization to any subsequent ICU readmissions, with discharges or deaths treated as censoring events. Following the approach of \cite{purushotham2018benchmarking}, we extract 26 clinical features for analysis. Our study aims to identify critical features strongly associated with ICU  readmission frequency. Understanding these associations offers valuable insights for healthcare providers, enabling more effective treatment planning, targeted monitoring, and resource allocation. In particular, identifying critical predictors of frequent readmission can support more proactive care strategies, potentially reducing ICU demand. Moreover, the results can aid in identifying high-risk populations and informing the development of more effective preventive healthcare policies.

The data processing workflow comprises several steps. First, we restrict our analysis to  patients admitted under the ``Emergency'' category, using covariates recorded during each patient's initial ICU admission. We then exclude the variables ``Min PaO2/FiO2,'' ``Min Bilirubin,'' and ``Max Bilirubin'' due to substantial missingness (exceeding 10,000 observations). To avoid collinearity, we retain only the ``Max'' records for these variables, resulting in a final set of 20 covariates for analysis. To mitigate the influence of extreme values in the clinical data, we apply truncation based on the distribution of each variable. For approximately symmetric variables, values beyond three standard deviations from the mean are truncated. For right-skewed variables, we truncate observations exceeding the 99th percentile. All continuous variables are then standardized to improve estimation stability. Additional filtering is applied to include only patients with event times ranging from 7.8 hours to six months. This lower bound removes approximately 180 censoring events shorter than 7.8 hours, which are treated as outliers. After preprocessing, the final dataset consists of 20,360 events from 18,252 patients. 

To fit the linear transformation model, it is crucial to identify a significant continuous variable to serve as the primary predictor, with its regression coefficient fixed at 1 for identifiability. To this end, we first fit a Cox-type recurrent event model using cubic B-splines with eight interior knots. The significant covariates are summarized in Table \ref{mimic_cox_table}, where EST represents the estimated coefficients and ESE denotes the estimated standard errors, calculated by inverting the covariance matrix obtained via the resampling method. According to Table \ref{mimic_cox_table}, the variable ``Max Blood Pressure'' exhibits the smallest p-value and has a negative coefficient. This result suggests that patients with lower blood pressure levels are at a higher risk of near-term ICU readmission. This finding is consistent with existing clinical literature. For instance, \cite{khanna2023association} examined the relationship between blood pressure components and organ dysfunction in septic patients, concluding that lower blood pressure (hypotension) is associated with an increased risk of ICU mortality.
\begin{table}[ht]
\centering
\caption{Estimated regression coefficients for the MIMIC-III data, using the Cox-type recurrent event model}
\begin{tabular}{l|lll|l}
\toprule[1.2pt]
Variables & EST  &  ESE  &   95\% CI    &   p-value    \\
\hline
Max Blood Pressure$^*$     & -0.0801  & 0.0227 &[-0.0356  ,  -0.1245]  &$<$ 0.001 \\
Min Sodium    & -0.0652& 0.0305   & [-0.1251 ,  -0.0054]  &  0.031 \\
Hematologic Malignancy  & 0.1992 &    0.1263     &[0.1145, 0.6096] &  0.001  \\
\toprule[1.2pt]
\end{tabular}
\label{mimic_cox_table}
\end{table}
We flip the sign of the ``Maximum Blood Pressure'' variable to align with its negative estimated coefficient, designate it as $X_1$, and impose the identifiability constraints $\beta_1=1$ and $\alpha(0.023)=1$, as 0.023 is close to the median of all observed event times. We then fit the linear transformation model using quartic B-splines, allowing the number of polynomial pieces to vary from 5 to 11. The optimal spline configuration is selected based on the Bayesian Information Criterion (BIC). The model with the lowest BIC contains 6 spline pieces for $\log (\alpha(t))$ and 11 spline pieces for $\log (q(u))$. We set the resampling size for estimating the covariance matrix to 5000. The significant variables identified under this model are summarized in Table \ref{mimic_ltm_table}.
\begin{table}[ht]
\begin{center}
\caption{Estimated parameters for the MIMIC-III data with the linear transformation model}
\begin{tabular}{l|lll|l}
\toprule[1.2pt]
Variables & EST  &  ESE  &   95\% CI    &   p-value    \\
\hline
Max Heartrate     & 0.1991 & 0.0541 &[0.0930, 0.3052]  &0.0002         \\
Min Heartrate     & -0.1301 & 0.0446  &[-0.2176,  -0.0426]  & 0.0036\\
Age & 0.1992 & 0.0523 & [0.0967,  0.3016] & 0.0001\\
Min Blood Pressure  &0.4243 &  0.0371   & [0.3515, 0.4970] & $<$0.0001 \\
Max Body Temperature & 0.0926 & 0.0327 &[0.0285,  0.1566] &0.0046 \\
Min Glasgow Coma Scale  &-0.0950 & 0.0312  &[-0.1562, -0.0338]& 0.0023\\
Admission Type & 0.2309 & 0.1161 & [0.0034,  0.4585]&0.0467\\
\toprule[1.2pt]
\end{tabular}
\label{mimic_ltm_table}
\end{center}
\end{table}

\begin{figure}[ht]
     \centering
     \begin{subfigure}[b]{0.3\textwidth}
         \centering
         \includegraphics[width=\textwidth]{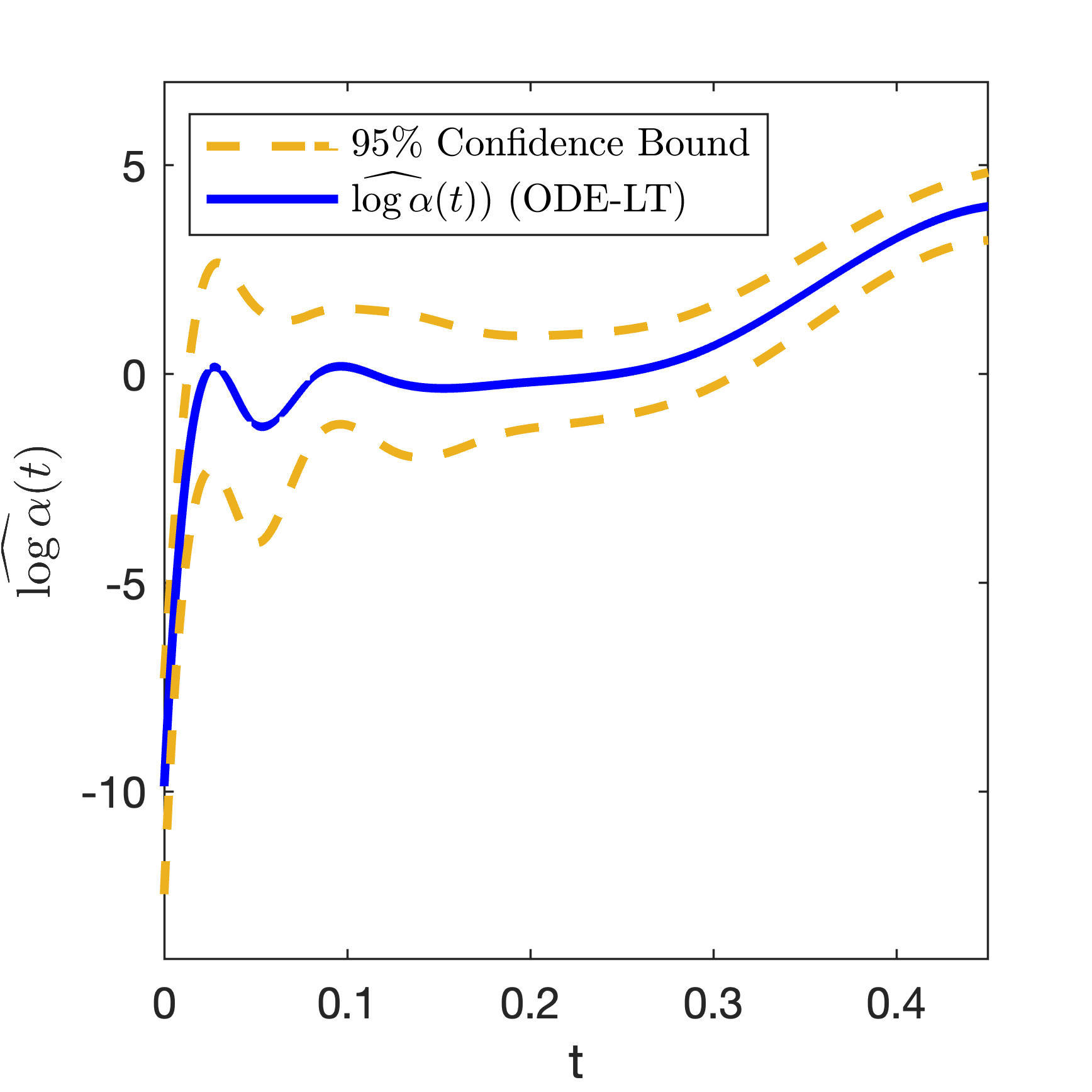}
         \caption{ $\widehat{\log \alpha}(t)$}
     \end{subfigure}
     \hfill
     \begin{subfigure}[b]{0.3\textwidth}
         \centering
         \includegraphics[width=\textwidth]{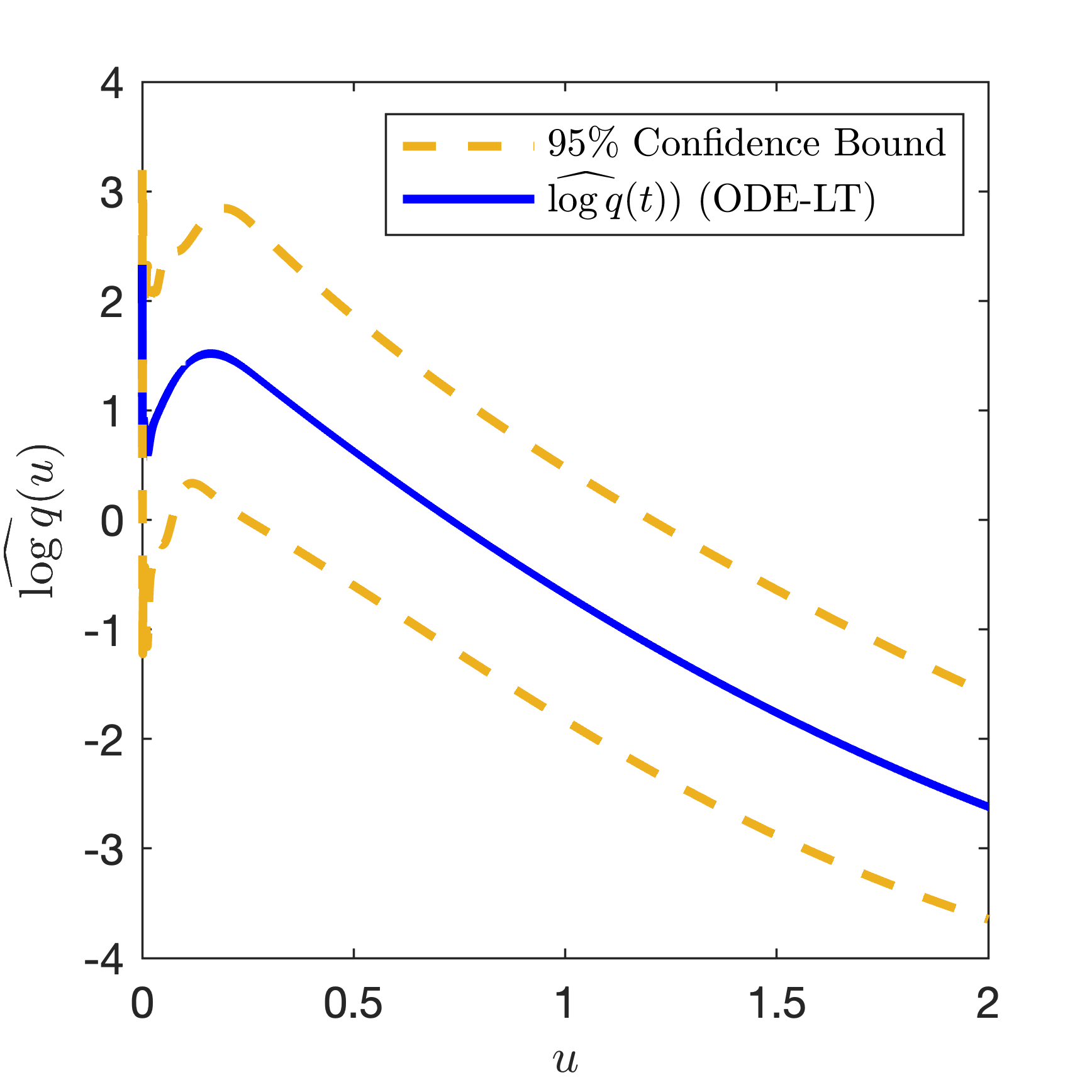}
         \caption{$\widehat{\log q}(u)$}
     \end{subfigure}
      \hfill
     \begin{subfigure}[b]{0.3\textwidth}
         \centering
         \includegraphics[width=\textwidth]{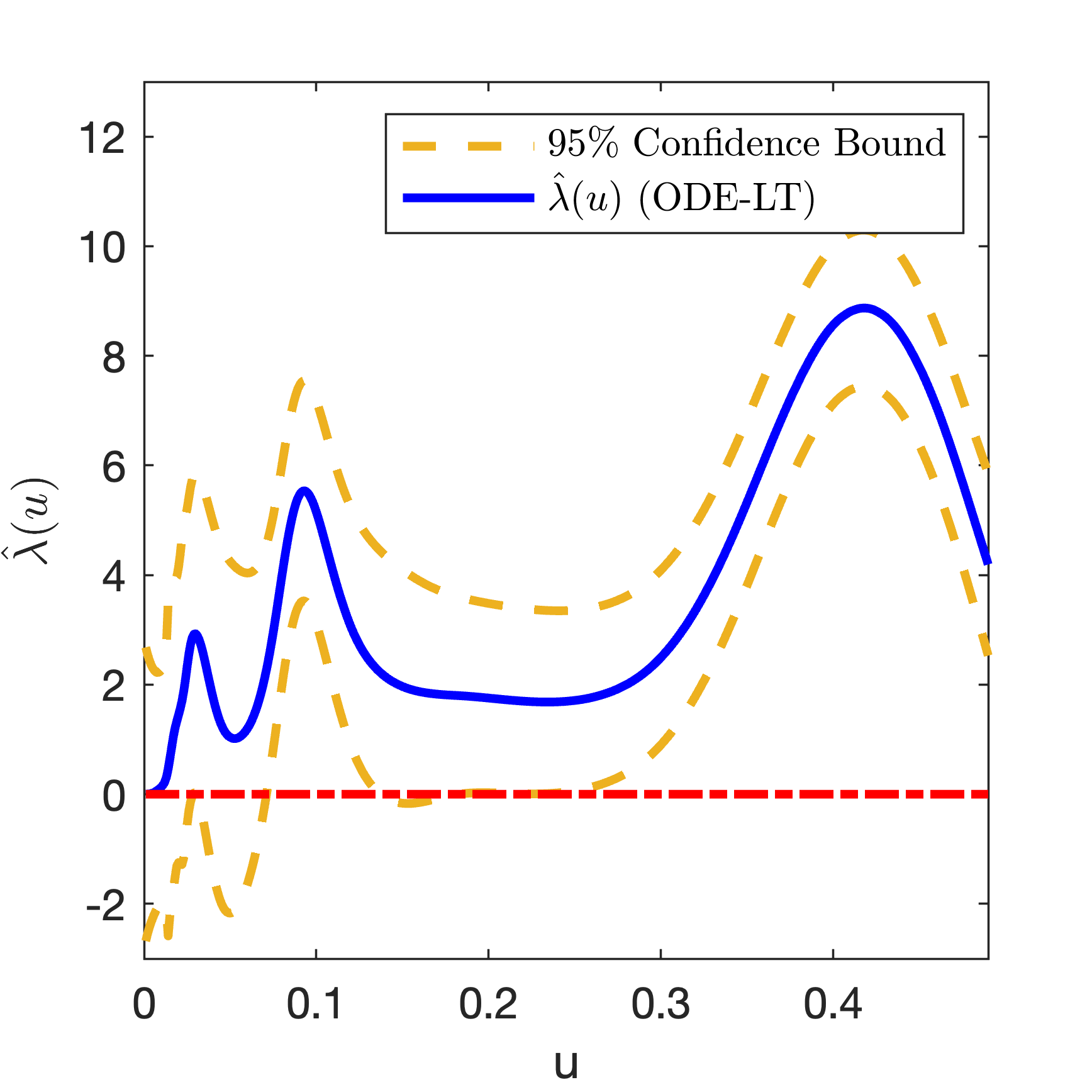}
         \caption{$\hat{\lambda}_{\Bar{x}}(t)$}
         \label{mimic_q}
     \end{subfigure}
     \caption{(a), (b): The spline estimators $\widehat{\log \alpha}(t)$ and $\widehat{\log q}(u)$ using the sieve MLE method for MIMIC-III data; (c) $\hat{\lambda}_{\Bar{x}}(t)$ conditional on the mean covariate $\Bar{x}$}
     \label{mimic_spline}
\end{figure}
Figure \ref{mimic_spline} presents the spline estimators $\widehat{\log\alpha}(t)$ and $\widehat{\log q}(u)$, along with their corresponding $95\%$ pointwise confidence intervals. Notably, neither $\log(\alpha(t))$ nor $\log (q(u))$ appears to be a constant function. Additional evidence, shown in Figure S1 in the Supplemental Materials, further indicates that neither $\log\alpha(t)$ nor $\log q(t)$ exhibits a linear relationship with $\log(t)$. Therefore, based on the identifiability conditions discussed in the Supplemental Materials, the fitted linear transformation model does not reduce to either the Cox-type or AFT-type models. We thus conclude that the general linear transformation model provides a better fit for the MIMIC-III dataset compared to the Cox-type and AFT-type alternatives.

As shown in Tables \ref{mimic_cox_table} and \ref{mimic_ltm_table}, the linear transformation model identifies more statistically significant variables than the Cox-type recurrent event model. Among these, ``Heart Rate'' emerges as a key predictor of ICU readmission frequency, with the estimated coefficients indicating that both excessively high maximum and exceedingly low minimum heart rates are associated with increased readmission risk. ``Age'' is another significant factor, with older patients exhibiting a higher likelihood of ICU readmission, consistent with general medical knowledge. The ``Glasgow Coma Scale (GCS)'', a widely used neurological assessment tool, also plays a key role; lower GCS scores, reflecting more severe neurological impairment, are linked to more frequent ICU readmissions. Additionally, the ``Admission Type'' variable, distinguishing surgical (1) from regular medical treatment (0) cases, is statistically significant, suggesting that patients undergoing surgery are more likely to experience recurrent ICU stays due to the severity of their conditions. It is important to note that our analysis is based solely on clinical data from patients' initial ICU admissions, which may diminish  predictive accuracy over time. Future research will focus on extending the model to incorporate time-varying covariates, thereby enhancing its predictive power and clinical relevance.

\section{Conclusion}\label{sec:conclusion}
In this paper, we propose an ordinary differential equation (ODE) framework that unifies many existing recurrent event models, including the Cox-type models \citep{andersen1982cox,pepe1993some}, the AFT-type models \citep{ghosh2003semiparametric}, and the linear transformation models \citep{zeng2006efficient, zeng2007}. Instead of modeling the intensity or hazard function as in \cite{zeng2007} and \cite{tang2023survival}, our method models the conditional mean function of the recurrent event process through the solution of an ODE, thereby avoiding the restrictive Poisson process assumption and allowing for more flexible dependence structures. Building  upon this framework, we develop a computationally efficient, gradient-based optimization algorithm for parameter estimation that leverages well-established ODE solvers and spline-based approximations. Moreover, for inference under the general linear transformation model, we introduce a resampling-based approach to estimate the covariance matrix of the parameter estimates. This method circumvents the need for complex density estimation procedures and is simple to implement in practice.
Notably, for commonly used models such as the Cox-type and AFT-type, the covariance matrix can also be estimated directly by solving the closed-form expression of matrix $A$ from Theorem 3, as detailed in the Supplemental Materials and demonstrated in our simulation studies.

Another interesting application of the proposed ODE framework is its ability to facilitate hypothesis testing to evaluate  whether a nested model adequately fits the observed recurrent event data. As discussed in the Supplemental Materials, if $\log(q(t))$ exhibits a linear relationship with $\log(t)$, it aligns with the Cox-type model. Similarly, a linear relation between $\log(\alpha(t))$ and $\log(t)$ corresponds to the AFT-type model. These characterizations allow for assessing the appropriateness of Cox-type or AFT-type models for the recurrent event data at hand by examining the estimated functional forms of $\log(q(t))$ and $\log(\alpha(t))$ plotted against $\log(t)$.

Beyond the aforementioned models, the proposed ODE framework enables the development of more flexible recurrent event model structures. For example, time-dependent covariates can be integrated into the ODE formulation to capture evovling individual-specific information over time. Moreover, the censoring times may, in practice, be influenced by a subject's  recurrent event process. To account for this dependency, one can introduce a shared unobservable frailty variable $Z$ to model the association between terminal and recurrent events. In such cases, the terminal and recurrent events may be described by a pair of coupled or independent ODEs, analogous to the joint modeling approaches explored in \cite{xu2017joint, xu2020generalized}.
Furthermore, this ODE framework can also be used to model the recurrent event process for the regression analysis of longitudinal data \citep[e.g.,][]{sun2005semiparametric, zhao2011semiparametric}.

\bibliography{./reference.bib}

\end{document}